\documentclass[english,prd,superscriptaddress,nofootinbib,preprintnumbers,eqsecnum]{revtex4}

\usepackage{amsmath}
\usepackage[latin1]{inputenc}
\usepackage{graphicx}
\usepackage{amssymb}
\usepackage{bm}
\usepackage{color}
\usepackage{amsfonts}
\usepackage{dcolumn}
\usepackage{color}
\usepackage{babel}

\def\be{\begin{equation}}
\def\ee{\end{equation}}
\def\ba{\begin{eqnarray}}
\def\ea{\end{eqnarray}}
\def\bs{\begin{subequations}}
\def\es{\end{subequations}}

\newcommand{\s}{{\cal S}}

\newcommand{\Z}{{\cal Z}}

\makeatother

\begin{document}

\title{Observational signatures of the theories beyond Horndeski}

\author{Antonio De Felice}
\affiliation{Yukawa Institute for Theoretical Physics, Kyoto University, 606-8502, Kyoto, Japan}

\author{Kazuya Koyama}
\affiliation{Institute of Cosmology \& Gravitation, University of Portsmouth, 
Portsmouth PO1 3FX, UK}

\author{Shinji Tsujikawa}
\affiliation{Department of Physics, Faculty of Science, Tokyo University of Science, 1-3,
Kagurazaka, Shinjuku, Tokyo 162-8601, Japan}

\date{\today}

\begin{abstract}

In the approach of the effective field theory of modified gravity, we derive the 
equations of motion for linear perturbations in the presence of 
a barotropic perfect fluid on the flat isotropic cosmological  
background. In a simple version of Gleyzes-Langlois-Piazza-Vernizzi (GLPV) theories, 
which is the minimum extension of Horndeski theories, we show that 
a slight deviation of the tensor propagation speed squared $c_{\rm t}^2$ from 1 generally 
leads to the large modification to the propagation speed squared $c_{\rm s}^2$ of a
scalar degree of freedom $\phi$. 
This problem persists whenever the kinetic energy $\rho_X$ of the field 
$\phi$ is much smaller than the background energy density $\rho_m$, 
which is the case for most of dark energy models in the asymptotic past. 
Since the scaling solution characterized by the constant ratio $\rho_X/\rho_m$ is 
one way out for avoiding such a problem, we study the evolution of perturbations 
for a scaling dark energy model in the framework of GLPV theories in 
the Jordan frame. 
Provided the oscillating mode of scalar perturbations is fine-tuned 
so that it is initially suppressed, 
the anisotropic parameter $\eta=-\Phi/\Psi$ 
between the two gravitational potentials $\Psi$ and $\Phi$ 
significantly deviates from 1 for $c_{\rm t}^2$ away from 1.
For other general initial conditions, the deviation of $c_{\rm t}^2$ from 1 
gives rise to the large oscillation of $\Psi$ with the frequency related 
to $c_{\rm s}^2$.
In both cases, the model can leave distinct imprints for 
the observations of CMB and weak lensing.

\end{abstract}

\maketitle

\section{Introduction}
\label{sec1} 

The constantly accumulating observational evidence for the late-time 
acceleration of the Universe \cite{SNIa,WMAP,BAO,Planck} implies that 
there may be at least one additional degree of freedom to the system 
of the Einstein-Hilbert action with non-relativistic matter and radiation. 
One simple example is a canonical scalar field $\phi$ 
with a sufficiently flat potential $V(\phi)$-- dubbed quintessence \cite{quin}.
The $\Lambda$-Cold-Dark-Matter ($\Lambda$CDM) model corresponds 
to the non-propagating limit of quintessence (i.e., the vanishing kinetic energy) 
with the constant potential $\Lambda$.

The scalar degree of freedom also arises in modified gravitational theories 
as a result of the breaking of gauge symmetries present
in General Relativity (GR) \cite{moreview}. 
In $f(R)$ gravity, for example, the presence of 
non-linear terms in the 4-dimensional Ricci scalar $R$ gives rise to the 
propagation of an extra gravitational scalar degree of 
freedom dubbed scalarons \cite{Staro,fRearly,fRreview}. 
Provided that the functional form of $f(R)$ is well 
designed \cite{fRviable}, it is possible 
to realize the late-time cosmic acceleration, while suppressing the 
propagation of the fifth force in regions of the high density through 
the chameleon mechanism \cite{chameleon}. 

Another well-known example of single-scalar modified gravitational theories 
is the covariant Galileon \cite{Galileons}, in which the field derivatives 
have couplings with the Ricci scalar $R$ and the Einstein tensor $G_{\mu \nu}$ 
(see Ref.~\cite{Nicolis} for the original Minkowski Galileon).
In this case the field kinetic terms drive the cosmic acceleration \cite{Sami,DT10}, 
while recovering the General Relativistic behavior in local regions 
through the Vainshtein mechanism \cite{Vain}.

Many dark energy models proposed in the literature (including 
quintessence, $f(R)$ gravity, and covariant Galileons)
can be accommodated in Horndeski theories \cite{Horndeski}--most 
general  second-order scalar-tensor theories with 
a single scalar field $\phi$ (see also Refs.~\cite{Deffayet}). 
Using the linear perturbation equations of motion on 
the flat Friedmann-Lema\^{i}tre-Robertson-Walker (FLRW) 
background \cite{DKT}, the dark energy models 
in the framework of Horndeski theories can be confronted 
with the observations of large-scale structure, CMB, 
and weak lensing \cite{Hoobser}.
In such general theories the screening mechanisms of 
the fifth force in local regions were also studied 
in Refs.~\cite{screening}.

There is another approach to the unified description of modified 
gravity-- the effective field theory (EFT) of cosmological 
perturbations \cite{Cheung}-\cite{Gao1}. 
By means of the Arnowitt-Deser-Misner (ADM) formalism 
with the 3+1 decomposition of space-time \cite{ADM}, 
one can construct several geometric 
scalar quantities from the extrinsic curvature 
$K_{\mu \nu}$ and the 3-dimensional 
intrinsic curvature ${\cal R}_{\mu \nu}$, e.g., 
$K \equiv {K^{\mu}}_{\mu}$, 
$\s \equiv K_{\mu \nu} K^{\mu \nu}$, 
${\cal R} \equiv {{\cal R}^{\mu}}_{\mu}$. 
The EFT of modified gravity is based on the expansion of 
a general Lagrangian $L$ in unitary gauge that depends on 
these geometric scalars, the lapse $N$, and the time $t$. 
In fact, Horndeski theories can be encompassed in such 
a general framework with several conditions imposed 
for the elimination of spatial derivatives higher 
than second order \cite{Piazza}.

Recently, Gleyzes {\it et al.} \cite{GLPV} proposed a generalized version 
of Horndeski theories by extending the Horndeski Lagrangian 
in the ADM form such that two additional constraints are not imposed.
For example the Horndeski Lagrangian involves 
the term $L_4=A_4(K^2-{\cal S})+B_4{\cal R}$, where 
the functions $A_4$ and $B_4$, which depend on $\phi$ and 
$X=\partial_{\mu}\phi\partial^{\mu}\phi$, 
have a particular relation $A_4=2X(\partial B_4/\partial X)-B_4$. 
In GR the ADM decomposition of the Einstein-Hilbert term $(M_{\rm pl}^2/2)R$, 
where $M_{\rm pl}$ is the reduced Planck mass,
leads to the Lagrangian $L_4$ with $B_4=-A_4=M_{\rm pl}^2/2$.
The theories with $B_4 \neq -A_4$ belong to a class of GLPV theories.

On the flat FLRW background, the Hamiltonian analysis based on linear 
cosmological perturbations shows that GLPV theories have only 
one scalar propagating degree of freedom \cite{GLPV,Lin,GlHa,Gao2}.
One distinguished feature of GLPV theories is that the scalar and matter 
sound speeds are coupled to each other \cite{Gergely,GLPV}. 
For example, in the covariantized version of the Minkowski Galileon where 
partial derivatives in the Lagrangian are replaced by covariant derivatives, 
the scalar propagation speed squared $c_{\rm s}^2$ becomes 
negative in the matter-dominated epoch due to 
a non-trivial kinetic-type coupling \cite{Kasecs}. 
Unlike the covariant Galileon with positive $c_{\rm s}^2$ during the matter era,
the covariantized Galileon mentioned above (a class of GLPV 
theories) is practically excluded as a viable dark energy scenario.

In this paper, we develop the analysis of cosmological perturbations further 
to confront dark energy models in GLPV theories with observations.
First, the linear perturbation equations of motion are derived 
in the presence of a barotropic perfect fluid for a general Lagrangian 
encompassing GLPV theories. 
We provide a convenient analytic formula for $c_{\rm s}^2$ and 
show that even a slight deviation from Horndeski theories generally 
gives rise to a non-negligible modification to the scalar sound speed.

We also apply our general formalism to a simple dark energy model 
with a canonical scalar field $\phi$ in which the function $B_4$ 
differs from $-A_4=M_{\rm pl}^2/2$. 
In this case the tensor propagation speed squared 
$c_{\rm t}^2=-B_4/A_4$ is different from 1. 
We show that this deviation leads to a significant modification to 
$c_{\rm s}^2$ whenever the field kinetic energy $\rho_X$ is suppressed 
relative to the background energy density $\rho_m$.
The scaling solution characterized by the constant $\rho_X/\rho_m$ 
is a possible way out to avoid having large values of $c_{\rm s}^2$ 
in the early cosmological epoch.

For the scaling dark energy model described by the potential 
$V(\phi)=V_1e^{-\lambda_1 \phi/M_{\rm pl}}+
V_2e^{-\lambda_2 \phi/M_{\rm pl}}$
($\lambda_1 \gtrsim 10$ and $\lambda_2 \lesssim 1$), we study 
the evolution of cosmological perturbations and resulting 
observational consequences.
For the initial conditions where the contribution of the oscillating mode 
$V_m^{(h)}$ to the velocity potential $V_m$ is 
suppressed, the evolution of perturbations 
is analytically known during the scaling matter era. 
In particular, the anisotropic parameter $\eta=-\Phi/\Psi$ between 
the two gravitational potentials $\Psi$ and $\Phi$ exhibits a large 
deviation from 1 for $c_{\rm t}^2$ away from 1.
If the oscillating mode $V_m^{(h)}$ gives a non-negligible contribution to 
$V_m$ initially, the rapid oscillations with frequencies related 
to $c_{\rm s}^2$ arise for the perturbations like $\Psi$ and $V_m$.
Thus the model in the framework of GLPV 
theories can be clearly distinguished from that in Horndeski theories.

This paper is organized as follows.
In Sec.~\ref{sec2} the extension of Horndeski theories to GLPV theories 
is briefly reviewed.
In Sec.~\ref{sec3} the perturbation equations of motion 
are derived in the presence of a barotropic perfect fluid 
according to the EFT approach encompassing GLPV theories.
In Sec.~\ref{speedsec} we obtain convenient formulae for the matter 
and scalar propagation speeds in the small-scale limit.
In Sec.~\ref{obsersec} we present observables associated with 
the measurements of large-scale structure, CMB, weak lensing, 
and discuss the quasi-static approximation on sub-horizon scales.
In Sec.~\ref{consec} we propose a simple dark energy model in 
the framework of GLPV theories and study its observational signatures by carefully 
paying attention to the oscillating mode induced by $c_{\rm s}^2$.
Sec.~\ref{concludesec} is devoted to conclusions.

\section{The theories beyond Horndeski}
\label{sec2} 

The EFT of cosmological perturbations is a powerful framework to deal with 
low-energy degrees of freedom in a systematic and unified way 
for a wide variety of modified gravity theories. 
It is based upon the 3+1 ADM decomposition of space-time 
described by the line element \cite{ADM}
\be
ds^{2}=g_{\mu \nu }dx^{\mu }dx^{\nu}
=-N^{2}dt^{2}+h_{ij}(dx^{i}+N^{i}dt)(dx^{j}+N^{j}dt)\,,  
\label{ADMmetric}
\ee
where $N$ is the lapse function, $N^i$ is the shift vector, and
$h_{ij}$ is the three-dimensional spatial metric. 
The extrinsic curvature is defined by  
$K_{\mu \nu}=h^{\lambda}_{\mu} n_{\nu;\lambda}$, 
where $n_{\mu}=(-N,0,0,0)$ is a unit vector orthogonal to 
the constant $t$ hyper-surfaces $\Sigma_t$ and a semicolon 
represents a covariant derivative.
We also introduce the three-dimensional Ricci tensor 
${\cal R}_{\mu \nu}={}^{(3)}R_{\mu \nu}$ on $\Sigma_t$. 
Then we can construct a number of geometric 
scalar quantities:
\be
K \equiv {K^{\mu}}_{\mu}\,,\qquad
\s \equiv K_{\mu \nu} K^{\mu \nu}\,,\qquad
{\cal R} \equiv
{{\cal R}^{\mu}}_{\mu}\,,\qquad
\Z \equiv {\cal R}_{\mu \nu}
\mathcal{R}^{\mu \nu}\,, \qquad
{\cal U} \equiv {\cal R}_{\mu \nu} K^{\mu \nu}\,. 
\label{ADMscalar}
\ee

Horndeski theories \cite{Horndeski} are the most general scalar-tensor theories 
with second-order equations of motion in generic space-time. 
The action of Horndeski theories is given by 
$S=\int d^4 x \sqrt{-g}\,L$ with the Lagrangian \cite{Deffayet}
\ba
L &=& G_2(\phi,X)+G_{3}(\phi,X)\square\phi 
+G_{4}(\phi,X)\, R-2G_{4,X}(\phi,X)\left[ (\square \phi)^{2}
-\phi^{;\mu \nu }\phi _{;\mu \nu} \right] \nonumber \\
& &
+G_{5}(\phi,X)G_{\mu \nu }\phi ^{;\mu \nu}
+\frac{1}{3}G_{5,X}(\phi,X)
[ (\square \phi )^{3}-3(\square \phi )\,\phi _{;\mu \nu }\phi ^{;\mu
\nu }+2\phi _{;\mu \nu }\phi ^{;\mu \sigma }{\phi ^{;\nu}}_{;\sigma}]\,,
\label{LH}
\ea
where $\square \phi \equiv (g^{\mu \nu} \phi_{;\nu})_{;\mu}$, 
and the four functions $G_{i}$ ($i=2,3,4,5$) depend on a scalar
field $\phi$ and its kinetic energy 
$X=g^{\mu \nu}\partial_{\mu} \phi \partial_{\nu} \phi$, 
with $G_{i,X} \equiv \partial G_i/\partial X$.
Choosing the unitary gauge $\phi=\phi(t)$ on the flat FLRW 
background described by the line element 
$ds^2=-dt^2+a^2(t)\delta_{ij}dx^i dx^j$, 
the Lagrangian (\ref{LH}) can be expressed 
in terms of the geometric scalars introduced above, as \cite{Piazza}
\be
L =
A_2(N,t)+A_3(N,t)K+A_4(N,t) (K^2-{\cal S}) 
+B_4(N,t){\cal R} +A_5(N,t) K_3 
+B_5(N,t) \left( {\cal U}-K {\cal R}/2 \right)\,,
\label{LGLPV}
\ee
where 
$K_3 \equiv K^3-3KK_{\mu \nu}K^{\mu \nu}
+2K_{\mu \nu}K^{\mu \lambda}{K^{\nu}}_{\lambda}$. 
Up to quadratic order in the perturbations we have 
$K_3=3H (2H^2-2KH+K^2-{\cal S})$, where 
$H=\dot{a}/a$ is the Hubble parameter (a dot 
represents a derivative with respect to $t$).

Horndeski theories satisfy the following 
two conditions \cite{Piazza}
\be
A_4=2XB_{4,X}-B_4\,,\qquad
A_5=-\frac13 XB_{5,X}\,.
\label{ABcon}
\ee
More concretely, the coefficients $G_i$ in Eq.~(\ref{LH}) 
and $A_i,B_i$ in Eq.~(\ref{LGLPV}) are 
related with each other, as
\ba
& & A_2=G_2-XF_{3,\phi}\,,\qquad
A_3=2(-X)^{3/2}F_{3,X}-2\sqrt{-X}G_{4,\phi}\,,\label{A3}
\nonumber \\
& & A_4=-G_4+2XG_{4,X}+XG_{5,\phi}/2\,,\qquad
B_4=G_4+X(G_{5,\phi}-F_{5,\phi})/2\,,\label{B4} 
\nonumber \\
& & A_5=-(-X)^{3/2}G_{5,X}/3\,,\qquad
B_5=-\sqrt{-X}F_{5}\,,\label{B5}
\label{AB}
\ea
where $F_3$ and $F_5$ are auxiliary functions obeying
the relations $G_3=F_3+2XF_{3,X}$ and $G_{5,X}=F_5/(2X)+F_{5,X}$. 
Since $X=-\dot{\phi}^2(t)/N^2$ in unitary gauge, 
the functional dependence of $\phi$ and $X$ can translate to 
that of $t$ and $N$.

It is possible to go beyond the Horndeski domain without imposing 
the two conditions (\ref{ABcon}) \cite{GLPV}.
This generally gives rise to derivatives higher than second order, but 
it does not necessarily mean that an extra propagating degree 
of freedom is present. In fact, the Hamiltonian analysis on the 
flat FLRW background shows that the theories described by 
the Lagrangian (\ref{LGLPV}), dubbed GLPV theories, 
do not possess an extra scalar mode of 
the propagation \cite{GLPV,Lin,GlHa,Gao2}.

\section{Perturbation equations of motion}
\label{sec3} 

The Lagrangian (\ref{LGLPV}) depends on $N, K, {\cal S}, {\cal R}, {\cal U}$, $t$, 
but not on ${\cal Z}$. The dependence on ${\cal Z}$ appears in the theories 
with spatial derivatives higher than second order \cite{Gao1,KaseIJMPD}, 
e.g., in Ho\v{r}ava-Lifshitz gravity \cite{Horava}. 
In the following we shall focus on the theories described 
by the action 
\be
S=\int d^4 x \sqrt{-g}\,L(N,K, {\cal S}, {\cal R}, {\cal U};t)+
\int d^4 x \sqrt{-g}\,L_m (g_{\mu \nu}, \Psi_m)\,,
\label{geneac}
\ee
where $L_m$ is the Lagrangian of the matter field $\Psi_m$. 
We consider a metric frame in which the scalar field $\phi$ 
is not directly coupled to matter (dubbed the Jordan frame).
For the matter component we consider a scalar field 
$\chi$ characterized by 
\be
L_m=P(Y)\,,\qquad Y=g^{\mu \nu} \partial_{\mu} \chi
\partial_{\nu} \chi\,,
\ee
whose description is the same as that of the barotropic 
perfect fluid \cite{Hu,Arroja,Mukoh}. 
The perfect fluids of radiation and non-relativistic matter can be 
modeled by $P(Y)=c_1 Y^2$ and 
$P(Y)=c_2(Y-Y_0)^2$ with $|(Y-Y_0)/Y_0| \ll 1$, respectively, 
where $c_1, c_2, Y_0$ are constants \cite{Scherrer,Kasecs}.

The linearly perturbed line element on the flat FLRW 
background with four metric perturbations 
$A, \psi, \zeta, E$ and tensor perturbations 
$\gamma_{ij}$ is given by \cite{Bardeen}
\be
ds^2=-(1+2A)dt^2+2 \partial_i \psi dt dx^i
+a^2(t) \left[ (1+2\zeta) \delta_{ij}
+2\partial_i \partial_j E+\gamma_{ij} \right]dx^i dx^j\,.
\label{permet}
\ee
Then the shift vector $N_i$ is related to the perturbation 
$\psi$, as 
\be
N_i=\partial_i \psi\,.
\ee
In the following we choose the gauge conditions 
\be
\delta \phi=0\,,\qquad
E=0\,,
\ee
under which temporal and spatial components of 
the gauge-transformation vector $\xi^{\mu}$ are fixed.
The background values (denoted by an overbar) 
of the ADM geometric quantities are
\be
\bar{K}_{\mu \nu}=H\bar{h}_{\mu \nu}\,,
\qquad \bar{K}=3H\,,\qquad 
\bar{\cal S}=3H^{2}\,,\qquad \bar{{\cal R}}_{\mu \nu}=0\,,
\qquad \bar{{\cal R}}=\bar{\cal U}=0\,.
\ee

Around the FLRW background we consider
the scalar perturbations $\delta N=N-1$, 
$\delta K_{\mu \nu}=K_{\mu \nu}-Hh_{\mu \nu}$, 
$\delta {\cal S}=2H\delta K+\delta
{K^{\mu}}_{\nu} {\delta K^{\nu}}_{\mu}$, and 
${\cal R}=\delta_1 {\cal R}+\delta_2 {\cal R}$, where 
$\delta_1 {\cal R}$ and $\delta_2 {\cal R}$ are 
the first-order and second-order perturbations respectively.
The scalar ${\cal U}$, which is a perturbed quantity itself, 
obeys the relation $\int d^4 x \sqrt{-g}\,\alpha(t) \,{\cal U}
=\int d^4 x \sqrt{-g} [\alpha(t) {\cal R} K/2+
\dot{\alpha}(t){\cal R}/(2N)]$ up to a boundary term,
where $\alpha(t)$ is an arbitrary function 
with respect to $t$.

Decomposing the scalar field $\chi$ as 
$\chi=\bar{\chi}(t)+\delta \chi(t, {\bm x})$ and 
omitting the overbar in the following discussion,  
the kinetic term $Y$, expanded up to second order, 
can be expressed in the form
$Y=-\dot{\chi}^2+\delta_1 Y+\delta_2 Y$, where
\ba
\delta_1 Y &=& 2\dot{\chi}^2 \delta N-2\dot{\chi} \dot{\delta \chi}\,,
\label{del1Y} \\
\delta_2 Y &=& -\dot{\delta \chi}^2-3\dot{\chi}^2 \delta N^2
+4\dot{\chi} \dot{\delta \chi} \delta N
+\frac{2\dot{\chi}}{a^2}\delta^{ij} \partial_i \psi \partial_j \delta \chi
+\frac{1}{a^2} (\partial \delta \chi)^2\,,
\label{delY}
\ea
and $(\partial \delta \chi)^2 \equiv \delta^{ij} 
\partial_i \delta \chi \partial_j \delta \chi$. 
The energy-momentum tensor of the field $\chi$ is given by 
$T_{\mu \nu}=Pg_{\mu \nu}-2P_{,Y} \partial_{\mu}\chi \partial_{\nu}\chi$. 
Defining the linear perturbations of energy density, momentum, and 
pressure, respectively, as 
$\delta T^0_0=-\delta \rho$, 
$\delta T^0_i=\partial_i \delta q$, and
$\delta T^i_j=\delta P \delta^i_j$, it follows that 
\be
\delta \rho=\left( P_{,Y}+2YP_{,YY} \right) \delta_1 Y\,,\qquad
\delta q=2P_{,Y} \dot{\chi} \delta \chi\,,\qquad 
\delta P=P_{,Y} \delta_1 Y\,.
\label{delrhoP}
\ee
These quantities appear in the perturbation equations 
of motion presented later.

\subsection{Background equations of motion}

Expanding the action (\ref{geneac}) up to linear order in scalar 
perturbations, we obtain the first-order action 
$S^{(1)}=\int d^4 x\,{\cal L}_1$ with \cite{Piazza,Gergely}
\be
{\cal L}_1=a^3 \left( \bar{L}+L_{,N}-3H {\cal F}-\rho \right) 
\delta N+3\left( \bar{L}-\dot{\cal F}-3H{\cal F} +P \right)
a^2 \delta a-2a^3 P_{,Y} \dot{\chi} \dot{\delta \chi}
+a^3{\cal E} \delta_1 {\cal R}\,,
\label{L1}
\ee
where 
\ba
{\cal F} &\equiv& L_{,K}+2H L_{,{\cal S}}\,,\\
{\cal E} &\equiv& L_{,{\cal R}}+\frac12 \dot{L}_{,{\cal U}}
+\frac32 H L_{,{\cal U}}\,,\\
\rho &\equiv& 2YP_{,Y}-P\,.
\label{density}
\ea
The last term of Eq.~(\ref{L1}) is a total derivative 
irrelevant to the background dynamics.
Varying Eq.~(\ref{L1}) with respect to 
$\delta N$, $\delta a$, and $\delta \chi$, respectively, 
we obtain the background equations of motion
\ba
& &
\bar{L}+L_{,N}-3H {\cal F}=\rho\,,
\label{back1}\\
& &
\bar{L}-\dot{\cal F}-3H{\cal F}=-P\,,
\label{back2}\\
& &
\frac{d}{dt} \left( a^3 P_{,Y} \dot{\chi} \right)=0\,.
\label{back3}
\ea
These correspond to the Hamiltonian constraint, 
the momentum constraint, and the equation of motion 
for $\chi$, respectively. 
Equation (\ref{back3}) is equivalent to the continuity 
equation $\dot{\rho}+3H(\rho+P)=0$ by using 
the definition (\ref{density}) of the field energy density.

\subsection{Perturbation equations of motion}

We expand the action (\ref{geneac}) up to second order 
in scalar perturbations to derive the linear perturbation 
equations of motion. 
In doing so, we use the following properties
\ba
\delta K^i_j &=&
\left( \dot{\zeta}-H \delta N \right) \delta^i_j
-\frac{1}{2a^2} \delta^{ik} \left( \partial_k N_j
+\partial_j N_k \right)\,,\\
\delta {\cal R}_{ij} &=&
-\left( \delta_{ij} \partial^2 \zeta+\partial_i
\partial_j \zeta \right)\,,
\ea
where $\partial^2 \zeta \equiv \delta^{ij} \partial_i \partial_j \zeta$. 
Expansion of the action (\ref{geneac}) gives rise to the terms 
in the forms $(\partial^2 \psi)^2/a^4$, 
$(\partial^2 \psi)(\partial^2 \zeta)/a^4$, and 
$(\partial^2 \zeta)^2/a^4$, which generate the spatial 
derivatives higher than second order. 
The absence of these higher-order terms 
requires that \cite{Piazza,Gergely}
\ba
& & L_{,KK}+4HL_{,{\cal S}K}+4H^2 L_{,{\cal SS}}+2L_{,{\cal S}}=0\,,
\label{con1}\\
& & L_{,K{\cal R}}+2HL_{,{\cal SR}}+\frac12 L_{,{\cal U}}
+HL_{,K{\cal U}}+2H^2 L_{,{\cal S U}}=0\,,\\
& & L_{,{\cal RR}}+2HL_{,{\cal RU}}+H^2 L_{,{\cal UU}}=0\,.
\label{con3}
\ea
The GLPV Lagrangian (\ref{LGLPV}) obeys all these conditions, 
so we assume the conditions (\ref{con1})-(\ref{con3}) 
in the following discussion.

The second-order Lagrangian density of 
${\cal L}_m=\sqrt{-g}\,P(Y)=N\sqrt{h}\,P(Y)$ reads
\be
{\cal L}_m^{(2)}=-\rho\, \delta N \delta \sqrt{h}-2P_{,Y} \dot{\chi} 
\dot{\delta \chi} \delta \sqrt{h}
+a^3 \left( P_{,Y} \delta_2 Y+\frac12 P_{,YY} \delta_1 Y^2
+P_{,Y} \delta N \delta_1 Y \right)\,,
\label{Lm2}
\ee
where $\delta \sqrt{h}=3a^3 \zeta$.
The Lagrangian density ${\cal L}=\sqrt{-g}\,L$ contains the second-order 
contribution $(\bar{L}+L_{,N}-3H {\cal F})\delta N \delta \sqrt{h}$. 
On using the background Eq.~(\ref{back1}), this cancels 
the first term of Eq.~(\ref{Lm2}). 
Then, expansion of the total action (\ref{geneac}) leads to 
the second-order action $S^{(2)}=\int d^4 x\, {\cal L}_{\rm T}^{(2)}$ with
\ba
{\cal L}_{\rm T}^{(2)}&=&
a^3 \biggl[ \left\{ L_{,N}+\frac12 L_{,NN}
-3H ({\cal W}-2L_{,{\cal S}}H) \right\} \delta N^2
+\left\{ {\cal W} \left( 3\dot{\zeta}-\frac{\partial^2 \psi}{a^2} \right)
-4({\cal D}+{\cal E}) \frac{\partial^2 \zeta}{a^2} \right\} \delta N
+4L_{,{\cal S}}\,\dot{\zeta} \frac{\partial^2 \psi}{a^2} \nonumber \\
& &-6L_{,{\cal S}} \dot{\zeta}^2+2{\cal E} \frac{(\partial \zeta)^2}{a^2}
-(P_{,Y}+2YP_{,YY})(\dot{\chi}\,\delta N-\dot{\delta \chi})^2
-6P_{,Y} \dot{\chi}\,\dot{\delta \chi}\,\zeta
-2P_{,Y}\dot{\chi}\,\delta \chi \frac{\partial^2 \psi}{a^2}
+P_{,Y} \frac{(\partial \delta \chi)^2}{a^2} \biggr]\,,
\label{LT2}
\ea
where
\ba
{\cal W} &\equiv& L_{,KN}+2HL_{,{\cal S}N}+4L_{,{\cal S}}H\,,\\
{\cal D} &\equiv& L_{,N{\cal R}}-\frac12 \dot{L}_{,{\cal U}}
+HL_{,N{\cal U}}\,.
\label{WD}
\ea

Varying Eq.~(\ref{LT2}) with respect to $\delta N$, 
$\partial^2 \psi$, $\zeta$, and $\delta \chi$ respectively, 
and employing Eqs.~(\ref{delrhoP}) and (\ref{back3}),
we obtain the following linear perturbation equations of motion 
\ba
& &
\left( 2L_{,N}+L_{,NN}-6H{\cal W}+12L_{,{\cal S}}H^2 \right) \delta N
+\left( 3\dot{\zeta}-\frac{\partial^2 \psi}{a^2} \right){\cal W}
-4({\cal D}+{\cal E})\frac{\partial^2 \zeta}{a^2}=\delta \rho\,,
\label{pereq1} \\
& &
{\cal W} \delta N-4L_{,{\cal S}} \dot{\zeta}=-\delta q\,,
\label{pereq2} \\
& &
\frac{1}{a^3} \frac{d}{dt} \left( a^3 {\cal Y} \right)
+4({\cal D}+{\cal E}) \frac{\partial^2 \delta N}{a^2}
+\frac{4{\cal E}}{a^2} \partial^2 \zeta+6P_{,Y} \dot{\chi}^2 \delta N
=3\delta P\,,
\label{pereq3} \\
& &
\frac{1}{a^3} \frac{d}{dt} \left[ a^3 (P_{,Y}+2YP_{,YY})
(\dot{\delta \chi}-\dot{\chi} \delta N) \right]
+3P_{,Y} \dot{\chi} \dot{\zeta} 
-\dot{\chi}P_{,Y} \frac{\partial^2 \psi}{a^2}
-P_{,Y} \frac{\partial^2 \delta \chi}{a^2}=0\,,
\label{pereq4}
\ea
where 
\be
{\cal Y} \equiv 4L_{,{\cal S}} \frac{\partial^2 \psi}{a^2}
-3\delta q\,.
\label{Xdef}
\ee

On using Eq.~(\ref{back3}), it is easy to show that the momentum 
perturbation $\delta q=2P_{,Y} \dot{\chi} \delta \chi$ obeys
\be
\dot{\delta q}+3H \delta q=-(\rho+P)\delta N-\delta P\,.
\label{mper1}
\ee
Similarly, Eq.~(\ref{pereq4}) can be expressed in the following form
\be
\dot{\delta \rho}+3H \left( \delta \rho+\delta P \right)
=-(\rho+P) \left( 3\dot{\zeta}-\frac{\partial^2 \psi}{a^2} \right)
-\frac{\partial^2 \delta q}{a^2}\,.
\label{mper2}
\ee
We note that Eqs.~(\ref{mper1}) and (\ref{mper2}) also follow from 
the continuity equations ${\delta T^{\mu}}_{i;\mu}=0$ and 
${\delta T^{\mu}}_{0;\mu}=0$, respectively.
Substituting Eq.~(\ref{Xdef}) into Eq.~(\ref{pereq3}) and using 
Eq.~(\ref{mper1}), it follows that 
\be
( \dot{L}_{,\cal S}+HL_{,\cal S} )\psi
+L_{,\cal S} \dot{\psi}+( {\cal D}+{\cal E} ) \delta N
+{\cal E} \zeta=0\,,
\label{pereq5}
\ee
where the integration constant is set to 0.
The dynamics of scalar perturbations is known by solving 
Eqs.~(\ref{pereq1}), (\ref{pereq2}), (\ref{pereq5}) together 
with Eqs.~(\ref{mper1}) and (\ref{mper2}).

For the tensor perturbation $\gamma_{ij}$ the second-order 
action reads $S_{h}^{(2)}=\int d^4 x\, {\cal L}_{h}^{(2)}$, 
where \cite{Piazza,Tsuji15,DT15} 
\be
{\cal L}_{h}^{(2)}=a^3 \frac{L_{,\cal S}}{4}
\delta^{ik} \delta^{jl} 
\left( \dot{\gamma}_{ij} \dot{\gamma}_{kl}
-\frac{c_{\rm t}^2}{a^2} \partial \gamma_{ij}
 \partial \gamma_{kl} \right)\,.
\ee
Here, the propagation speed squared is
\be
c_{\rm t}^2= \frac{{\cal E}}{L_{,\cal S}}\,.
\label{ct}
\ee
Then the equation of motion for gravitational waves 
is given by
\be
\ddot{\gamma}_{ij}+\left( 3H+\frac{\dot{L_{,\cal S}}}{L_{,\cal S}}
\right) \dot{\gamma}_{ij}-c_{\rm t}^2 \frac{\partial^2 \gamma_{ij}}
{a^2}=0\,.
\label{GWeq}
\ee
Provided that $L_{,\cal S}>0$ and $c_{\rm t}^2>0$, 
the ghost and Laplacian instabilities are absent 
for the tensor mode.

\section{Propagation speeds of scalar perturbations}
\label{speedsec}

We derive the propagation speeds of the gravitational scalar 
and the matter field in the small-scale limit.
In doing so, we first express $\delta N$ and $\partial^2 \psi/a^2$ 
in terms of $\zeta$, $\delta \chi$ and their derivatives by using
Eqs.~(\ref{pereq1}) and (\ref{pereq2}).
Substituting these relations into Eq.~(\ref{LT2}), 
the Lagrangian density can be expressed 
in the form \cite{Gergely,Kasecs}
\be
\mathcal{L}_{2}=a^{3}\left( \dot{\vec{\mathcal{X}}}^{t}{\bm K} 
\dot{\vec{\mathcal{X}}}-\partial _{j}\vec{\mathcal{X}}^{t}{\bm G}
\partial^{j}{\vec{\mathcal{X}}}-\vec{\mathcal{X}}^{t}{\bm B} 
\dot{\vec{\mathcal{X}}}-\vec{\mathcal{X}}^{t}{\bm M} \vec{\mathcal{X}}\right) \,,
\label{L2mat}
\ee
where ${\bm K}$, ${\bm G}$, ${\bm B}$, ${\bm M}$ 
are $2 \times 2$ matrices and 
$\vec{\mathcal{X}}^{t}=\left( \zeta, \delta \chi /M_{\mathrm{pl}} 
\right)$. The components of the two matrices 
${\bm K}$ and ${\bm G}$ are given, respectively, by 
\ba
& &
K_{11}=Q_s+\frac{16L_{,\s}^2}{M_{\rm pl}^2{\cal W}^2} 
\dot{\chi}^2 K_{22}\,,\qquad
K_{22}=(2\dot{\chi}^2P_{,YY}-P_{,Y})M_{\rm pl}^2\,,
\qquad
K_{12}=K_{21}=-\frac{4L_{,\s}\dot{\chi}}{M_{\rm pl}{\cal W}}
K_{22}\,,\nonumber \\
& & 
G_{11}=2(\dot{\cal M}+H{\cal M}-{\cal E})\,,\qquad
G_{22}=-P_{,Y}M_{\rm pl}^2\,,\qquad
G_{12}=G_{21}=-\frac{{\cal M}\dot{\chi}}{L_{,\s}M_{\rm pl}}G_{22}\,,
\ea
where 
\ba
Q_s &\equiv& 6L_{,\cal S}+\frac{8L_{,\cal S}^2}{{\cal W}^2}
(2L_{,N}+L_{,NN}-6H{\cal W}+12H^2L_{,\cal S})\,,\\
{\cal M} &\equiv& \frac{4L_{,\cal S}({\cal D}+{\cal E})}{{\cal W}}\,.
\ea

The scalar ghosts are absent as long as the determinants of principal 
sub-matrices of ${\bm K}$ are positive, which translates to the 
two conditions $K_{11}>0$ and $Q_sK_{22}>0$. These conditions 
are satisfied for $Q_s>0$ and $K_{22}>0$.

The dispersion relation following from Eq.~(\ref{L2mat})
in the limit of the large wave number $k$ is given by 
${\rm det} \left( \omega^2 {\bm K}-k^2{\bm G}/a^2 
\right)=0$. The scalar propagation speed $c_s$, 
which is related to the frequency $\omega$ as
$\omega^2=c_s^2\,k^2/a^2$ obeys
\be
\left(c_s^2 K_{11}-G_{11} \right)
\left(c_s^2 K_{22}-G_{22} \right)
-\left(c_s^2 K_{12}-G_{12} \right)^2=0\,.
\label{csso}
\ee
In Horndeski theories there is the specific relation 
${\cal D}+{\cal E}=L_{,\cal S}$ and hence
$G_{12}/K_{12}=G_{22}/K_{22}$. 
In this case the two solutions to $c_s^2$ 
are given by 
\ba
c_m^2 &=&\frac{G_{22}}{K_{22}}=
\frac{P_{,Y}}{P_{,Y}-2\dot{\chi}^2 P_{,YY}}
=\frac{\delta P}{\delta \rho}\,,
\label{cm} \\
c_{\rm H}^2 &=& \frac{1}{Q_s} \left[ 
G_{11}-(K_{11}-Q_s)\frac{G_{22}}{K_{22}} \right]
=\frac{2}{Q_s} \left( \dot{\cal M}+H{\cal M}-{\cal E} 
+\frac{8L_{,\cal S}^2\dot{\chi}^2 P_{,Y}}{{\cal W}^2} \right)\,.
\label{csHo}
\ea

In GLPV theories the relation ${\cal D}+{\cal E}=L_{,\cal S}$ no longer 
holds, so we define the parameter \cite{GlHa}
\be
\alpha_{\rm H} \equiv \frac{{\cal D}+{\cal E}}{L_{,\cal S}}-1
=c_{\rm t}^2-1+\frac{{\cal D}}{L_{,\cal S}}\,,
\label{alphaH}
\ee
which characterizes the deviation from Horndeski theories. 
We also introduce the following quantity 
\be
\beta_{\rm H} \equiv 2c_m^2 
\left( \frac{K_{11}}{Q_s}-1\right)\alpha_{\rm H}
=\frac{16L_{,\cal S}^2 (\rho+P)}{{\cal W}^2 Q_s}\alpha_{\rm H}\,.
\label{betaH}
\ee
If ${\cal D}=0$, then the parameter $\alpha_{\rm H}$ is 
simply related to the deviation of the tensor propagation speed 
squared from 1, as $\alpha_{\rm H}=c_{\rm t}^2-1$. 
This is the case for the theories with $B_4=B_4(t)$ 
and constant $B_5$. In Sec.~\ref{consec} we shall discuss the dynamics 
of cosmological perturbations for a simple model 
satisfying the condition ${\cal D}=0$.

Eliminating the terms $G_{22}$, $G_{11}$, $G_{12}$, and $K_{11}$ 
in Eq.~(\ref{csso}) with the help of Eqs.~(\ref{cm}), (\ref{csHo}), (\ref{betaH}) 
and the relation $G_{12}/K_{12}=(1+\alpha_{\rm H})G_{22}/K_{22}$, 
the two solutions to Eq.~(\ref{csso}) can be expressed as 
\ba
\tilde{c}_m^2 &=& 
\frac12 \left[ c_m^2+c_{\rm H}^2-\beta_{\rm H}
+ \sqrt{(c_m^2-c_{\rm H}^2+\beta_{\rm H})^2+
2c_m^2 \alpha_{\rm H} \beta_{\rm H}} \right]\,,
\label{cmexact} \\
c_{\rm s}^2 &=&
\frac12 \left[ c_m^2+c_{\rm H}^2-\beta_{\rm H}
- \sqrt{(c_m^2-c_{\rm H}^2+\beta_{\rm H})^2+
2c_m^2 \alpha_{\rm H} \beta_{\rm H}} \right]\,.
\label{csexact}
\ea
For non-relativistic matter with $c_m^2=0$, 
Eqs.~(\ref{cmexact}) and (\ref{csexact}) reduce to $\tilde{c}_m^2=0$ 
and $c_{\rm s}^2=c_{\rm H}^2-\beta_{\rm H}^2$ respectively.

For the general perfect fluid with $c_m^2 \neq 0$, we consider the case 
in which the deviation from Horndeski theories 
is small, i.e., $|\alpha_{\rm H}| \ll 1$.
Then the propagation speeds (\ref{cmexact}) and (\ref{csexact}) 
are approximately given, respectively, by 
\ba
\tilde{c}_m^2 &\simeq& 
c_m^2-\frac{c_m^2}
{2(c_{\rm H}^2-c_m^2-\beta_{\rm H})} \alpha_{\rm H} \beta_{\rm H}\,,
\label{cmap} \\
c_{\rm s}^2 &\simeq& 
c_{\rm H}^2-\beta_{\rm H}+\frac{c_m^2}
{2(c_{\rm H}^2-c_m^2-\beta_{\rm H})} \alpha_{\rm H} \beta_{\rm H}\,.
\label{csap}
\ea
The condition, $|\alpha_{\rm H}| \ll 1$, does not necessarily mean that 
$|\beta_{\rm H}|$ is also much smaller than 1.
In the covariantized Galileon model \cite{GLPV} where the partial derivatives 
of the original Minkowski Galileon \cite{Nicolis} are replaced by the covariant 
derivatives, we have $\beta_{\rm H}=3/10$ 
and $c_{\rm H}^2=11/40$ 
during the matter era for late-time tracking solutions 
(in which regime $|\alpha_{\rm H}|$ is much smaller than 1) \cite{Kasecs}. 
The reason why $|\beta_{\rm H}|$ is not as small as $|\alpha_{\rm H}|$ 
comes from the fact that the variable $Q_s/M_{\rm pl}^2$ in Eq.~(\ref{betaH}) is 
much smaller than 1 in the early cosmological epoch. 
Since $c_{\rm s}^2=-1/40$ in this case, the covariantized Galileon is 
plagued by the Laplacian instability on small scales.
On the other hand, for the covariant Galileon \cite{Deffayet}, 
we have $c_{\rm s}^2=c_{\rm H}^2=1/40$ 
during the matter era \cite{DT10}.

Generally, the sound speed squared $c_{\rm s}^2$ is subject to the 
modification arising from the
deviation from Horndeski theories, such that 
$c_{\rm s}^2 \simeq c_{\rm H}^2-\beta_{\rm H}$. 
Meanwhile, provided that $|\alpha_{\rm H} \beta_{\rm H}| \ll 1$,
the correction to the matter sound speed squared $c_m^2$, i.e.,  
the second term on the r.h.s.\ 
of Eq.~(\ref{cmap}), is suppressed to be small. 
This shows that the effect beyond Horndeski 
theories arises for the scalar sound speed $c_{\rm s}$ 
rather than the matter sound speed $\tilde{c}_{m}$.
In Sec.~\ref{consec} we shall apply the results in this section to 
a concrete model that belongs to a class of GLPV theories.

\section{Confrontations with observations}
\label{obsersec} 

In this section we discuss several physical quantities associated with 
the measurements of large-scale structures, CMB, and weak lensing 
in order to confront GLPV theories with observations.
We then proceed to the discussion of the quasi-static 
approximation for the perturbations deep inside the 
sound horizon.

\subsection{Observables}

We first introduce the gauge-invariant combinations of the matter 
density contrast $\delta_m$ and the velocity perturbation $v_m$, as
\be
\delta_m \equiv \delta-3Hv\,,\qquad
v_m \equiv v+(1+w) \frac{\delta \phi}{\dot{\phi}}\,,
\label{delvm}
\ee
where 
\be
\delta \equiv \frac{\delta \rho}{\rho}\,,\qquad
v \equiv \frac{\delta q}{\rho}\,,\qquad
w \equiv \frac{P}{\rho}\,.
\label{delta}
\ee
Since we choose the unitary gauge ($\delta \phi=0$), 
the perturbation $v_m$ is equivalent to $v$ itself.
In Fourier space we can rewrite Eqs.~(\ref{mper1}) and 
(\ref{mper2}), respectively, as 
\ba
& &
\dot{v}_m+3H \left( c_m^2-w \right)v_m=
-(1+w)\delta N-c_m^2 \delta_m\,,
\label{mattereq1} \\
& &
\dot{\delta}_m+3(Hv_m)^{\cdot}
+3H  \left( c_m^2-w \right) 
 \left( \delta_m+3Hv_m \right)
 =-(1+w) \left( 3\dot{\zeta}
+\frac{k^2}{a^2}\psi \right)
+\frac{k^2}{a^2}v_m\,,
\label{mattereq2}
\ea
where $c_m$ is defined by Eq.~(\ref{cm}).

Since we are interested in the growth of structures 
after the onset of the matter era, we shall focus on the case of non-relativistic 
matter characterized by $w=0$ and $c_m^2=0$.  
Taking the time derivative of Eq.~(\ref{mattereq2}) and using 
Eq.~(\ref{mattereq1}), we obtain
\be
\ddot{\delta}_m+2H \dot{\delta}_m
+\frac{k^2}{a^2}\Psi=
-3\ddot{B}-6H\dot{B}\,, 
\label{mattereq}
\ee
where $B \equiv \zeta+Hv_m$, and $\Psi$ is the gauge-invariant 
gravitational potential defined by \cite{Bardeen}
\be
\Psi \equiv \delta N+\dot{\psi}\,.
\label{Psidef}
\ee
If $c_m$ is not exactly 0 and the term $-c_m^2 \delta_m$ on 
the r.h.s.\ of Eq.~(\ref{mattereq1}) is not neglected, this gives rise to the term 
$c_m^2(k^2/a^2)\delta_m$ on the l.h.s. of Eq.~(\ref{mattereq}).
This works as a pressure that prevents the gravitational growth 
induced by the source term $(k^2/a^2)\Psi$. 
The matter propagation speed squared $\tilde{c}_m^2$ 
in GLPV theories is not equivalent to $c_m^2$, but, in the 
limit $c_m^2 \to 0$, they are identical to each other.
On the other hand, the scalar sound speed squared $c_{\rm s}^2$ is 
generally subject to a non-negligible change even by the slight 
deviation from Horndeski theories.

In order to know the evolution of the matter density contrast $\delta_m$, 
we need to relate the gravitational potential $\Psi$ in (\ref{mattereq}) 
with $\delta_m$. Usually, this relation is expressed 
in the following form 
\be
\frac{k^2}{a^2}\Psi =-4\pi G_{\rm eff} \rho\,\delta_m\,,
\label{Psieq}
\ee
where $G_{\rm eff}$ is the effective gravitational coupling. 
In GR, $G_{\rm eff}$ is equivalent to the Newton gravitational 
constant $G$. In modified gravitational theories, $G_{\rm eff}$ 
generally differs from $G$. 
In Horndeski theories, for example, the quasi-static approximation 
provides the analytic expression of $G_{\rm eff}$ for the perturbations 
deep inside the sound horizon \cite{DKT}. 
Provided that the terms on the r.h.s.\ 
 of Eq.~(\ref{mattereq}) are 
negligible compared to those on the l.h.s., the evolution of 
$\delta_m$ is known by integrating Eq.~(\ref{mattereq}). 
The growth rate of $\delta_m$ is related to the peculiar velocity 
of galaxies \cite{Kaiser}. 
For the observations of redshift-space distortions, the quantity 
$f\sigma_8$ is usually introduced \cite{Tegmark}, where
\be
f \equiv \frac{\dot{\delta}_m}{H \delta_m}\,,
\ee
and $\sigma_8$ is 
the amplitude of over-density at the comoving $8\,h^{-1}$ Mpc 
scale ($h$ is the normalized Hubble constant 
$H_0=100 h$ km\,sec$^{-1}$ Mpc$^{-1}$). 

In order to confront modified gravity models with the observations of 
CMB and weak lensing, we also introduce
the following gauge-invariant gravitational potential 
\be
\Phi \equiv \zeta+H \psi\,.
\ee
The effective gravitational potential associated with the deviation of 
light rays is given by \cite{Sapone}
\be
\Phi_{\rm eff} \equiv \frac12 \left( \Psi-\Phi \right)\,.
\label{Phieff}
\ee
Introducing the anisotropic parameter
\be
\eta \equiv -\frac{\Phi}{\Psi}\,,
\label{etadef}
\ee
Eq.~(\ref{Phieff}) can be expressed as $\Phi_{\rm eff}=(1+\eta)\Psi/2$.
In GR we have $\eta=1$ and hence $\Phi_{\rm eff}=\Psi$.
We caution that the definition (\ref{etadef}) is valid only for $\Psi \neq 0$.
If the gravitational potential $\Psi$ crosses 0 with oscillations, 
we should compute $\Phi_{\rm eff}$ from Eq.~(\ref{Phieff}) 
rather than using $\eta$. 
As we will see in Sec.~\ref{consec}, the crossing of $\Psi=0$ 
can actually occur in GLPV theories if the oscillating mode 
initially dominates the perturbation $v_m$.

\subsection{The quasi-static approximation on sub-horizon scales}

For the observations of large-scale structure and weak lensing,  
we are primarily interested in the evolution of perturbations for the 
modes deep inside the sound horizon ($c_{\rm s}k \gg aH$). 
In the presence of a propagating scalar degree of freedom, 
there is an oscillating mode of the field perturbation in addition 
to the mode induced by matter perturbations.
Provided that the oscillating mode of perturbations  
is suppressed relative to the matter-induced mode, 
the time derivatives of metric perturbations 
(like $\dot{\zeta}$ and $\dot{\psi}$) can be neglected  
relative to the terms involving their spatial 
derivatives \cite{quasi}.
In Horndeski theories, this quasi-static approximation 
was first employed in Ref.~\cite{DKT} to derive 
the analytic expression of $G_{\rm eff}$, 
$\Phi_{\rm eff}$, and $\eta$.

In GLPV theories, let us discuss what kind of 
difference from Horndeski theories arises. 
First of all, Eq.~(\ref{pereq5}) can be written in the form
\be
(1+\alpha_{\rm H}) \Psi+c_{\rm t}^2 \zeta+ 
\left(1+\frac{ \dot{L}_{,\cal S}}{HL_{,\cal S}} \right)H\psi=
\alpha_{\rm H}\dot{\psi} \,.
\label{quasi2}
\ee
Since the time derivative $\alpha_{\rm H} \dot{\psi}$ 
does not vanish, we need to deal with Eq.~(\ref{quasi2}) as the differential 
equation rather than the constraint equation. 

Under the quasi-static approximation on sub-horizon scales, the dominant 
contributions to Eq.~(\ref{pereq1}) can be regarded as those involving 
the Laplacian terms $\partial^2 \psi/a^2$, $\partial^2 \zeta/a^2$ and  
$\delta \rho$. Then, in Fourier space, Eq.~(\ref{pereq1}) reduces to 
\be
{\cal W} \frac{k^2}{a^2} \psi+4L_{,{\cal S}} (1+\alpha_{\rm H})
\frac{k^2}{a^2} \zeta \simeq \rho \delta\,.
\label{quasi1}
\ee
Taking the time derivative of Eq.~(\ref{pereq1}) and 
eliminating the term $\dot{\delta \rho}$ on account of 
Eq.~(\ref{mper2}), the quasi-static approximation 
for sub-horizon perturbations leads to 
\be
{\cal W} \Psi+(\dot{\cal W}+H{\cal W}+\rho)\psi
+4\left[ (1+\alpha_{\rm H})(\dot{L}_{,\cal S}+HL_{,\cal S})
+L_{,\cal S} \dot{\alpha}_{\rm H} \right]\zeta 
\simeq -4\alpha_{\rm H}L_{,{\cal S}} \dot{\zeta}\,.
\label{quasi3}
\ee
In deriving Eq.~(\ref{quasi3}) we have implicitly assumed that 
the mass $m_{\phi}$ of the scalar degree of freedom $\phi$ 
is at most of the order of the Hubble parameter $H$. 
In some of the modified gravity models in which the chameleon 
mechanism is at work \cite{fRviable}, $m_{\phi}$ can be much larger 
than $H$ as we go back to the past. In the regime $m_{\phi} \gg H$, 
however, the scalar degree of freedom is nearly frozen, 
so that the evolution of perturbations is similar to 
that in GR \cite{DKT}. 
The modification of gravity manifests itself 
in the late cosmological epoch associated with the cosmic acceleration, 
in which regime $m_{\phi}$ is at most of the order of $H$.

In Horndeski theories we have $\alpha_{\rm H}=0$, so the terms 
on the r.h.s.\ 
of Eqs.~(\ref{quasi2}) and (\ref{quasi3}) identically vanish. 
In this case, we can express $\zeta$ and $\psi$ in terms of $\Psi$
by using Eqs.~(\ref{quasi2}) and (\ref{quasi3}). 
Substituting these relations into Eq.~(\ref{quasi1}), 
we obtain the modified Poisson equation (\ref{Psieq}) 
with the effective gravitational coupling $G_{\rm eff}$. 
The effective gravitational potential $\Phi_{\rm eff}$ 
and the anisotropic parameter $\eta$ are known 
accordingly. This procedure is given in Appendix A.

In GLPV theories the time derivatives $\dot{\psi}$ and 
$\dot{\zeta}$ are present, so we cannot derive
the closed-form expression of $G_{\rm eff}$ and 
$\Phi_{\rm eff}$. The existence of the time-derivative terms
in Eqs.~(\ref{quasi2}) and (\ref{quasi3}) implies that the 
oscillating mode may play a non-trivial role for the evolution 
of perturbations. 
In Sec.~\ref{consec} we shall discuss the condition 
under which the oscillating mode is suppressed relative 
to the matter-induced mode for a model in the framework 
of GLPV theories.

\section{A concrete model}
\label{consec} 

In this section we consider a concrete model in which 
Horndeski theories are minimally extended to GLPV theories.
The model is described by the action (\ref{geneac})  
with the Lagrangian 
\be
L=A_2+A_4 (K^2-{\cal S})+B_4 {\cal R}\,,
\label{concrete}
\ee
where 
\be
A_2=-\frac12 X-V(\phi)\,,\qquad A_4=-\frac12 M_{\rm pl}^2\,,\qquad
B_4=\frac12 M_{\rm pl}^2 F(\phi)\,,
\label{AB2}
\ee
with $V(\phi)$ and $F(\phi)$ being functions of $\phi$. 
The model with $F(\phi)=1$ correspond to GR in the presence of 
a minimally coupled scalar field $\phi$ with a potential $V(\phi)$.
If the function $F(\phi)$ differs from 1, then the first condition of 
Eq.~(\ref{ABcon}) is not satisfied, so this is already beyond 
the realm of Horndeski theories.
For the matter Lagrangian $L_m$ we consider a single 
perfect fluid described by the constant equation 
of state $w=P/\rho$.

{}From Eqs.~(\ref{back1})-(\ref{back2}) we obtain the background equations of motion 
\ba
& & 3M_{\rm pl}^2 H^2=\frac12 \dot{\phi}^2+V(\phi)+\rho\,,\label{ba1} \\
& & -2M_{\rm pl}^2 \dot{H}=\dot{\phi}^2+\rho+P\,. \label{ba2}
\ea
Taking the time derivative of Eq.~(\ref{ba1}) and using Eqs.~(\ref{back3}) 
and (\ref{ba2}), the scalar field $\phi$ obeys
\be
\ddot{\phi}+3H \dot{\phi}+V_{,\phi}=0\,.
\label{ba3}
\ee
Equations (\ref{ba1})-(\ref{ba3}) are equivalent to those in GR.
This means that, at the background level, we cannot distinguish between 
the theories with same $A_4$ but with different $B_4$ \cite{Kasecs}.

For the later convenience we introduce the following variables \cite{CLW}
\be
x_1 \equiv \frac{\dot{\phi}}{\sqrt{6}H M_{\rm pl}}\,,\qquad
x_2 \equiv \frac{\sqrt{V}}{\sqrt{3}HM_{\rm pl}}\,,\qquad 
\Omega_m \equiv \frac{\rho}{3M_{\rm pl}^2 H^2}\,,\qquad
\lambda \equiv -\frac{M_{\rm pl}V_{,\phi}}{V}\,,
\ee
with which the Friedmann equation (\ref{ba1}) can be written as 
$\Omega_m=1-x_1^2-x_2^2$. 
The dimensionless quantities $x_1$, $x_2$, and 
$x_3 \equiv \phi/M_{\rm pl}$ obey the equations of motion 
\ba
x_1' &=& -3x_1+\frac{\sqrt{6}}{2} \lambda x_2^2+\frac32 x_1 
\left[ (1-w)x_1^2+(1+w)(1-x_2^2) \right]\,,\\
x_2' &=& -\frac{\sqrt{6}}{2}\lambda x_1 x_2+\frac32 x_2 
\left[ (1-w)x_1^2+(1+w)(1-x_2^2) \right]\,,\\
x_3' &=& \sqrt{6}x_1\,,
\ea
where a prime represents a derivative with respect to 
${\cal N}=\ln a$. Note that the variable $\lambda$ 
is known as a function of $x_3$.

\subsection{Propagation speeds}

The background degeneracy between the theories with different 
values of $B_4$ is broken at the level of cosmological perturbations.
{}From Eq.~(\ref{ct}) the tensor propagation speed squared is given by 
\be
c_{\rm t}^2=-\frac{B_4}{A_4}=F(\phi)\,.
\ee
The deviation from Horndeski theories 
can be quantified by the difference of $c_{\rm t}^2$ from 1. 
In fact, the deviation parameter $\alpha_{\rm H}$ 
in Eq.~(\ref{alphaH}) reads
\be
\alpha_{\rm H}=c_{\rm t}^2-1\,.
\ee
The quantities $c_{\rm H}^2$ and $\beta_{\rm H}$ defined 
in Eqs.~(\ref{csHo}) and (\ref{betaH}) reduce, respectively, to 
\ba
c_{\rm H}^2 &=&
1+\frac{2M_{\rm pl}^2 H^2}
{\dot{\phi}^2} \left[ ( 1-c_{\rm t}^2) 
\frac{\dot{H}}{H^2}+\frac{2c_{\rm t}\dot{c}_{\rm t}}{H} \right]\,,\\
\beta_{\rm H} &=& 
2(1-c_{\rm t}^2) \left( 1+\frac{2M_{\rm pl}^2 \dot{H}}
{\dot{\phi}^2} \right)\,,
\ea
where we used the background Eq.~(\ref{ba2}). 
In the regime $|\alpha_{\rm H}| \ll 1$, the sound speed 
squared (\ref{csap}) can be estimated as 
$c_{\rm s}^2 \simeq c_{\rm H}^2-\beta_{\rm H}$, i.e.,  
\be
c_{\rm s}^2 \simeq 1-2(1-c_{\rm t}^2)
-\frac{2M_{\rm pl}^2 H^2}
{\dot{\phi}^2} \left[ (1-c_{\rm t}^2) 
\frac{\dot{H}}{H^2}-\frac{2c_{\rm t}\dot{c}_{\rm t}}{H} \right]\,.
\label{crms}
\ee

The density parameters of the field kinetic energy and the 
potential energy are given, respectively, by $\Omega_X=x_1^2$ and 
$\Omega_V=x_2^2$. In terms of these parameters, 
Eq.~(\ref{ba2}) can be expressed as 
\be
\frac{\dot{H}}{H^2}=-3\Omega_X-\frac32 (1+w) \Omega_m\,.
\ee
Then, the sound speed squared (\ref{crms}) reduces to 
\be
c_{\rm s}^2 \simeq 1-(c_{\rm t}^2-1) 
\left[ (1+w) \frac{\Omega_m}{2\Omega_X}-1 \right]
+\frac{2c_{\rm t} \dot{c}_{\rm t}}{3H\Omega_X}\,.
\label{cses}
\ee

If the field $\phi$ is responsible for dark energy, its kinetic energy 
is usually suppressed relative to the background energy 
density in the early cosmological epoch, i.e., 
$\Omega_X/\Omega_m \ll 1$ (unless we consider early 
dark energy models). 
For $c_{\rm t}^2>1$ the sound speed squared (\ref{cses}) 
can be negative, which leads to the Laplacian instability 
on small scales. 
Let us consider the case in which $\dot{c}_{\rm t}=0$, i.e., 
$F(\phi)=$\,constant.
In order to realize the condition $c_{\rm s}^2>0$ during the 
radiation and early matter eras, we require that 
\be
c_{\rm t}^2-1\, \lesssim\, \frac{\Omega_X}{\Omega_m}\,.
\label{ctcon}
\ee
As we go back to the past the ratio $\Omega_X/\Omega_m$ 
gets smaller, so $c_{\rm t}^2$ needs to be very close to 1 in the 
early cosmological era. 

When $c_{\rm t}^2<1$ the Laplacian instability is absent, 
but $c_{\rm s}$ becomes highly super-luminal. 
This means that the quantity $c_{\rm s}k/a$, which appears 
in the perturbation equations of motion, becomes much larger 
than $k/a$, so that the perturbation theory can break down 
for $c_{\rm s}k/a$ above some cut-off scale $M$.

We stress that the problem of large values of $|c_{\rm s}^2|$ persists 
whenever there is a scalar field $\phi$ whose kinetic energy 
is suppressed relative to the background energy density. 
The slowly varying scalar potentials $V(\phi)$ responsible 
for the late-time cosmic acceleration 
are generally plagued by this problem.

One way out is to consider a primordial scaling field characterized by 
the constant ratio $\Omega_X/\Omega_m$ \cite{Ratra,CLW,Joyce}. 
The scaling solution can be realized by the exponential potential 
$V(\phi)=V_1 e^{-\lambda_1 \phi/M_{\rm pl}}$ for the constant 
$\lambda_1$ satisfying $\lambda_1^2>3(1+w)$ \cite{CLW}.
For the success of the big bang nucleosynthesis, the slope 
$\lambda_1$ is constrained to be $\lambda_1>9.4$ \cite{Bean}. 
In this case the late-time 
cosmic acceleration is not realized, so the form of the potential 
needs to be modified to exit from the scaling regime. 
One of such models is given by \cite{Nelson}
\be
V(\phi)=V_1 e^{-\lambda_1 \phi/M_{\rm pl}}
+V_2 e^{-\lambda_2 \phi/M_{\rm pl}}\,,
\label{poten}
\ee
where $V_1, V_2, \lambda_1, \lambda_2$ are constants.
Provided that $\lambda_2^2<2$, the solutions finally approach 
the acceleration attractor characterized by the field equation 
of state $w_{\rm DE}=-1+\lambda_2^2/3$.

\subsection{Perturbation equations and the oscillating mode}

Since our primary interest is the growth of structures after the onset 
of the matter-dominated epoch, we consider a non-relativistic 
perfect fluid characterized by $w=0$ and $c_m^2=0$ 
for the matter Lagrangian $L_m=P(Y)$. For example, 
the Lagrangian $P(Y)=c_2(Y-Y_0)^2$ with $|(Y-Y_0)/Y_0| \ll 1$
can describe a non-relativistic perfect fluid \cite{Scherrer,Kasecs}. 
In the following we also focus on the case 
in which $c_{\rm t}^2$ is constant.

First, we introduce the following 
dimensionless quantities
\be
V_m \equiv H v_m\,,\qquad
\chi \equiv H \psi\,.
\ee
For the model (\ref{concrete}), the perturbation equations of motion 
in Fourier space following from Eqs.~(\ref{pereq1}), (\ref{pereq2}), and 
(\ref{mper1})-(\ref{pereq5}) are given by 
\ba
\zeta' &=& \delta N+\frac32 \Omega_m V_m\,,\label{zetaeq}\\
\chi' &=& \left( \frac{H'}{H}-1 \right) 
\chi-c_{\rm t}^2 \delta N-c_{\rm t}^2 \zeta\,,\label{tpsi}\\
\delta' &=& -\frac32 \Omega_m \delta +3(\Omega_X-1) 
\delta N+\left( \frac{k}{aH} \right)^2 (c_{\rm t}^2 \zeta+
V_m)\,,\label{delmde}\\
V_m' &=&
-\delta N+\frac{H'}{H} 
V_m\,,\label{vmde} \\
\Omega_X \delta N &=& \frac12 \Omega_m \delta_m
-\frac13 \left( \frac{k}{aH} \right)^2 
(c_{\rm t}^2 \zeta+\chi)\,,
\label{delN}
\ea
where $\delta_m=\delta-3V_m$ 
is the gauge-invariant matter perturbation.

The two gravitational potentials $\Psi$ and $\Phi$ are
given, respectively, by $\Psi=\delta N-(H'/H)\chi+\chi'$ 
and $\Phi=\zeta+\chi$.
{}From Eq.~(\ref{tpsi}) we obtain the following relation 
\be
\Psi+\Phi=(1-c_{\rm t}^2)(\delta N+\zeta)\,.
\label{ani}
\ee
Taking the ${\cal N}$ derivative of Eqs.~(\ref{delmde})-(\ref{vmde}) 
and using other equations of motion, it follows that 
\be
\delta_m^{''}+\left( 2+\frac{H'}{H} \right) \delta_m^{'}
-\frac32 \Omega_m \delta_m
=\left( \frac{k}{aH} \right)^2  (c_{\rm t}^2-1) \delta N
-12\,\Omega_X \delta N+18\,\Omega_X \left( 1+\frac{H'}{H}
+\frac{\Omega_X'}{2\Omega_X} 
\right) V_m\,.
\label{delms}
\ee
Since $\Psi=-\Phi$ for $c_{\rm t}^2=1$, the anisotropic 
parameter $\eta$ defined by Eq.~(\ref{etadef}) is equivalent to 1. 
In this case the first term on the r.h.s.\ 
of Eq.~(\ref{delms}) also vanishes. 
Provided $\Omega_X \ll 1$, the remaining terms 
on the r.h.s.\ 
of Eq.~(\ref{delms}) are suppressed relative to those on the l.h.s., 
so we obtain the growing-mode solution $\delta_m \propto a$ 
in the deep matter era ($H'/H \simeq -3/2$ and $\Omega_m \simeq 1$).

In GLPV theories, there is the anisotropic stress ($\eta \neq 1$) 
between the gravitational potentials induced by the difference 
of $c_{\rm t}^2$ from 1. 
Unlike the usual modified gravity models, the parameter
$\eta$ deviates from 1 even in the early cosmological epoch.
We also note that the existence of the first term on the 
r.h.s.\ 
of Eq.~(\ref{delms}) leads to the modified growth of 
matter perturbations relative to the case $c_{\rm t}^2=1$.

In order to understand the evolution of the velocity potential, we take 
the ${\cal N}$ derivative of Eq.~(\ref{vmde}) and then use other 
equations of motion. 
The resulting second-order equation for $V_m$ reads
\be
V_m''+\alpha_1 V_m'+\alpha_2 V_m=-\left( \frac{k}{aH} \right)^2\chi\,,
\label{Vmeq}
\ee
where 
\ba
\alpha_1 &\equiv& 
\frac{\sqrt{6}}{4x_1} \left[ 4\lambda (1-\Omega_m-x_1^2)
-\sqrt{6} x_1 (2-\Omega_m-2x_1^2) \right]\,,\label{alpha1} \\
\alpha_2 &\equiv&
-(c_{\rm t}^2-1)\frac{\Omega_m}{2x_1^2} \left( \frac{k}{aH} \right)^2
\nonumber \\
& &
+\frac{3}{2x_1} \left[ 3x_1 \{ 4x_1^4+2(2\Omega_m-3)x_1^2
+\Omega_m (\Omega_m-1) \}-\sqrt{6}\lambda 
(\Omega_m+4x_1^2)(\Omega_m+x_1^2-1) \right]\,.
\label{alpha2}
\ea
The general solution to Eq.~(\ref{Vmeq}) can be expressed 
in the following form
\be
V_m=V_{m}^{(s)}+V_{m}^{(h)}\,,
\ee
where $V_{m}^{(s)}$ 
is the special solution and
$V_{m}^{(h)}$ is the homogeneous solution derived by 
setting the r.h.s.\ 
 of Eq.~(\ref{Vmeq}) to be 0.
As long as the first term on the r.h.s.\ 
 of Eq.~(\ref{alpha2}) 
dominates over the other terms, the special solution 
is given by 
\be
V_{m}^{(s)} \simeq \frac{1}{c_{\rm t}^2-1} \frac{2\Omega_X}
{\Omega_m} \chi=-\frac{1}{c_{\rm s}^2-c_{\rm t}^2}\chi\,.
\label{Vms}
\ee
In the second equality we used the fact that, for 
$c_m^2=0$, the sound speed squared $c_{\rm s}^2$ 
is exactly given by Eq.~(\ref{cses}) with $w=0$.

The oscillations of perturbations are induced by 
the homogenous solution $V_{m}^{(h)}$. 
In order to see the behavior of oscillations, we consider 
the scaling solution during the matter-dominated epoch.
The scaling matter era can be realized by the dominance 
of the potential $V_1 e^{-\lambda_1 \phi/M_{\rm pl}}$ in 
Eq.~(\ref{poten}), which corresponds to \cite{CLW}
\be
x_1=x_2=\frac{\sqrt{6}}{2\lambda_1}\,,\qquad
\Omega_m=1-\frac{3}{\lambda_1^2}\,.
\label{x1scaling}
\ee
Substituting Eq.~(\ref{x1scaling}) into Eqs.~(\ref{alpha1})-(\ref{alpha2}) 
with $\lambda=\lambda_1$, 
we find that the homogenous solution satisfies
\be
{V_{m}^{(h)}}''+\frac32{V_{m}^{(h)}}'+
\frac13 (\lambda_1^2-3) \left[ (1-c_{\rm t}^2) 
\left( \frac{k}{aH} \right)^2+\frac{27}{2\lambda_1^2} 
\right]V_{m}^{(h)}=0\,.
\label{Vm}
\ee

During the scaling matter era the scale factor evolves as 
$a \propto t^{2/3}$, so the evolution of the quantity 
\be
K \equiv \frac{k}{aH}
\ee
is known as $K({\cal N})=K_i e^{{\cal N}/2}$, 
where $K_i$ is initial value of $K$ at ${\cal N}=0$. 
Then Eq.~(\ref{Vm}) can be expressed as 
\be
{U_{m}^{(h)}}''+\left[ c_{\rm eff}^2 K_i^2 e^{{\cal N}}
+\frac{9(7\lambda_1^2-24)}{16\lambda_1^2} \right] 
U_{m}^{(h)}=0\,,
\label{Umeq}
\ee
where 
\be
U_{m}^{(h)} \equiv a^{3/4} V_{m}^{(h)}\,,\qquad
c_{\rm eff}^2 \equiv \frac13 (\lambda_1^2-3) (1-c_{\rm t}^2)\,.
\ee
If $c_{\rm t}^2>1$, then we have $c_{\rm eff}^2<0$ and hence 
the perturbation $U_{m}^{(h)}$ is prone to the Laplacian 
instability on small scales. When $c_{\rm t}^2<1$,
the perturbation exhibits oscillations induced by the positive 
Laplacian term $c_{\rm eff}^2 K^2({\cal N})$.
Note that $c_{\rm eff}^2$ is related to $c_{\rm s}^2$ as
$c_{\rm eff}^2=c_{\rm s}^2-c_{\rm t}^2$. 
As long as $c_{\rm s}^2 \gg c_{\rm t}^2$, $c_{\rm eff}^2$ is 
approximately equivalent to $c_{\rm s}^2$.

The solution to Eq.~(\ref{Umeq}) in the regime $c_{\rm t}^2<1$
is given by 
\be
U_{m}^{(h)} = c_1 J_{\nu} (x)+c_2 Y_{\nu} (x)\,,
\label{Vmh}
\ee
where $c_1,c_2$ are integration constants, $J_{\nu}(x)$ and 
$Y_{\nu}(x)$ are the Bessel functions of first and second kinds 
respectively, with 
\be
\nu \equiv \frac{3\sqrt{7\lambda_1^2-24}}{2\lambda_1}i\,,\qquad
x \equiv 2c_{\rm eff} K({\cal N})=\frac{2c_{\rm eff}k}{aH}\,.
\ee

Provided that $x \gg 1$, the first term in the square bracket 
of Eq.~(\ref{Umeq}) is much larger than the second term (which is 
of the order of 1). In this case the solution to Eq.~(\ref{Umeq})
is given by Eq.~(\ref{Vmh}) with the index $\nu \simeq 0$. 
Using the asymptotic forms of the Bessel functions in the limit 
$x \gg 1$, we obtain the following approximate solution
\be
V_{m}^{(h)} \simeq a^{-3/4} \sqrt{\frac{2}{\pi x}}
\left[ c_1 \cos \left( x-\frac{\pi}{4} \right)+
c_2 \sin \left( x-\frac{\pi}{4} \right)
\right]\,.
\label{Vmh2}
\ee
The perturbation $V_{m}^{(h)}$ exhibits a 
damped oscillation with the frequency determined by 
$c_{\rm eff}k$. Since $x \propto e^{{\cal N}/2}$ in the scaling matter era, 
the amplitude decreases as $|V_{m}^{(h)}| \propto a^{-1}$.
As we will see later, the homogenous solution (\ref{Vms}) stays nearly 
constant during the matter era. 
This means that, as we go back to the past, the oscillating solution 
(\ref{Vmh2}) tends to dominate over the homogenous solution 
(as it happens for the dark energy models in $f(R)$ theories \cite{fRviable}).
The dominance of the oscillating solution can be avoided for
the initial conditions satisfying $|V_{m}^{(h)}| \ll |V_{m}^{(s)}|$. 
This amounts to choosing the initial conditions of $V_m$ close to 
the value (\ref{Vms}) and $V_m' \simeq 0$.

\subsection{Evolution of perturbations}

Numerically we integrate the perturbation equations of motion 
(\ref{zetaeq})-(\ref{vmde}) with (\ref{delN}) in order to 
find the precise evolution of $V_m$, $\delta_m$ 
as well as the gravitational potentials $\Psi$ and $\Phi$.

First, we study the case in which the special solution to 
Eq.~(\ref{Vmeq}) dominates over the homogenous 
solution (\ref{Vmh2}), notwithstanding that this 
requires fine-tuning of the initial conditions for the perturbations.
The initial conditions corresponding to such a case are given by 
$V_m \simeq -K^2 \chi/\alpha_2$ and $V_m' \simeq 0$. 
From Eq.~(\ref{vmde}) the latter condition implies that 
$\delta N \simeq (H'/H)V_m$, in which case the r.h.s.\ 
 of Eq.~(\ref{zetaeq}) is small such that $\zeta' \simeq 0$. 
As in the case of GR, we also employ the initial condition $\chi'=0$. 
Then, for given $\delta_m$, the initial conditions 
of $\zeta$, $\chi$, $V_m$, and $\delta$ are 
known accordingly.

\begin{figure}
\includegraphics[width=3.2in]{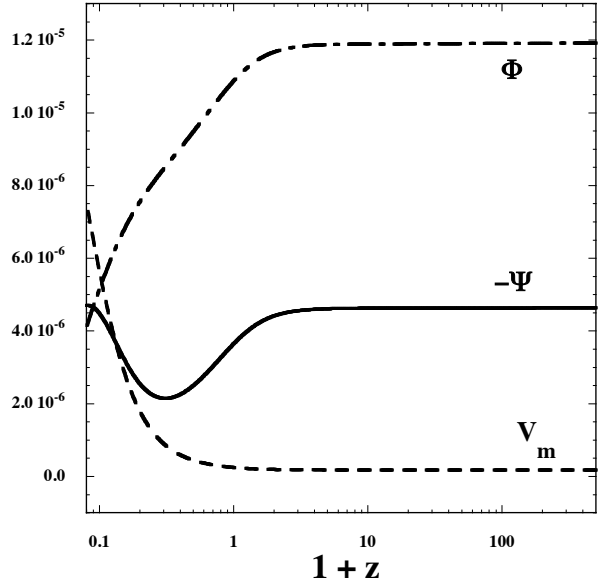}
\caption{
The evolution of the velocity potential $V_m$ and the gravitational 
potentials $-\Psi$ and $\Phi$ versus $1+z$~$(=a_0/a)$ 
for $c_{\rm t}^2=0.5$, $\lambda_1=10$, 
$\lambda_2=0$, and $V_2/V_1=10^{-6}$. 
The present epoch ($z=0$) is identified as $\Omega_m=0.3$.
We choose the initial conditions as
$x_1=x_2=\sqrt{6}/(2\lambda_1)$, 
$x_3=0$, $\zeta=1.4832 \times 10^{-5}$, 
$\chi=-2.9124 \times 10^{-6}$, 
$\delta=1.9351 \times 10^{-3}$, and $V_m=1.8007 \times 10^{-7}$
for the wave number $K=k/(aH)=25$ at the redshift $z=527.9$.
These correspond to the initial conditions with the negligible 
oscillating mode characterized by 
$V_m=-K^2\chi/\alpha_2$ and $V_m'=0$.
\label{fig1}}
\end{figure}

In Fig.~\ref{fig1} we plot the evolution of $V_m$, $-\Psi$, and 
$\Phi$ versus the redshift $z=a_0/a-1$ ($a_0$ is the today's value of $a$)
for $c_{\rm t}^2=0.5$, $\lambda_1=10$, 
$\lambda_2=0$, and $V_2/V_1=10^{-6}$.
During the matter-dominated epoch, the background solutions 
are in the scaling regime characterized by Eq.~(\ref{x1scaling}). 
Numerically we find that the perturbations $V_m$, $\zeta$, $\chi$, 
and $\delta N$ stay nearly constant by the end of the matter era
for the initial conditions explained above. 
On using the approximate relations (\ref{Vms}) as well as $V_m' \simeq 0$
and $\chi' \simeq 0$ with $H'/H=-3/2$, the evolution of perturbations 
during the scaling matter era can be estimated as
\ba
& &
V_m \simeq \frac{ \Omega_m \delta_m}
{c_{\rm s}^2 K^2-3\Omega_X}\,,\qquad
\delta N \simeq -\frac32 \frac{ \Omega_m \delta_m}
{c_{\rm s}^2 K^2-3\Omega_X}\,,\nonumber \\
& &
\zeta \simeq \frac{5c_{\rm s}^2-2c_{\rm t}^2}
{2c_{\rm t}^2 (c_{\rm s}^2K^2-3\Omega_X)}\Omega_m \delta_m\,,\qquad
\chi \simeq -\frac{c_{\rm s}^2-c_{\rm t}^2}
{c_{\rm s}^2K^2-3\Omega_X}\Omega_m \delta_m\,.
\label{anaso1}
\ea
Then the gauge-invariant gravitational potentials satisfy 
\be
\Psi \simeq -\frac{3(1+c_{\rm s}^2-c_{\rm t}^2)}
{2(c_{\rm s}^2 K^2-3\Omega_X)}\Omega_m \delta_m\,,\qquad
\Phi \simeq \frac{5c_{\rm s}^2-2c_{\rm t}^2(1+c_{\rm s}^2-c_{\rm t}^2)}
{2c_{\rm t}^2(c_{\rm s}^2 K^2-3\Omega_X)}
\Omega_m \delta_m\,,
\label{anaso2}
\ee
so that the anisotropic parameter is 
simply given by 
\be
\eta \simeq 1+\frac{5(1-c_{\rm t}^2)(c_{\rm s}^2-c_{\rm t}^2)}
{3c_{\rm t}^2(1+c_{\rm s}^2-c_{\rm t}^2)}\,,
\label{etaana}
\ee
where $c_{\rm s}^2=c_{\rm t}^2+(\lambda_1^2-3)(1-c_{\rm t}^2)/3$.

\begin{figure}
\includegraphics[width=3.5in]{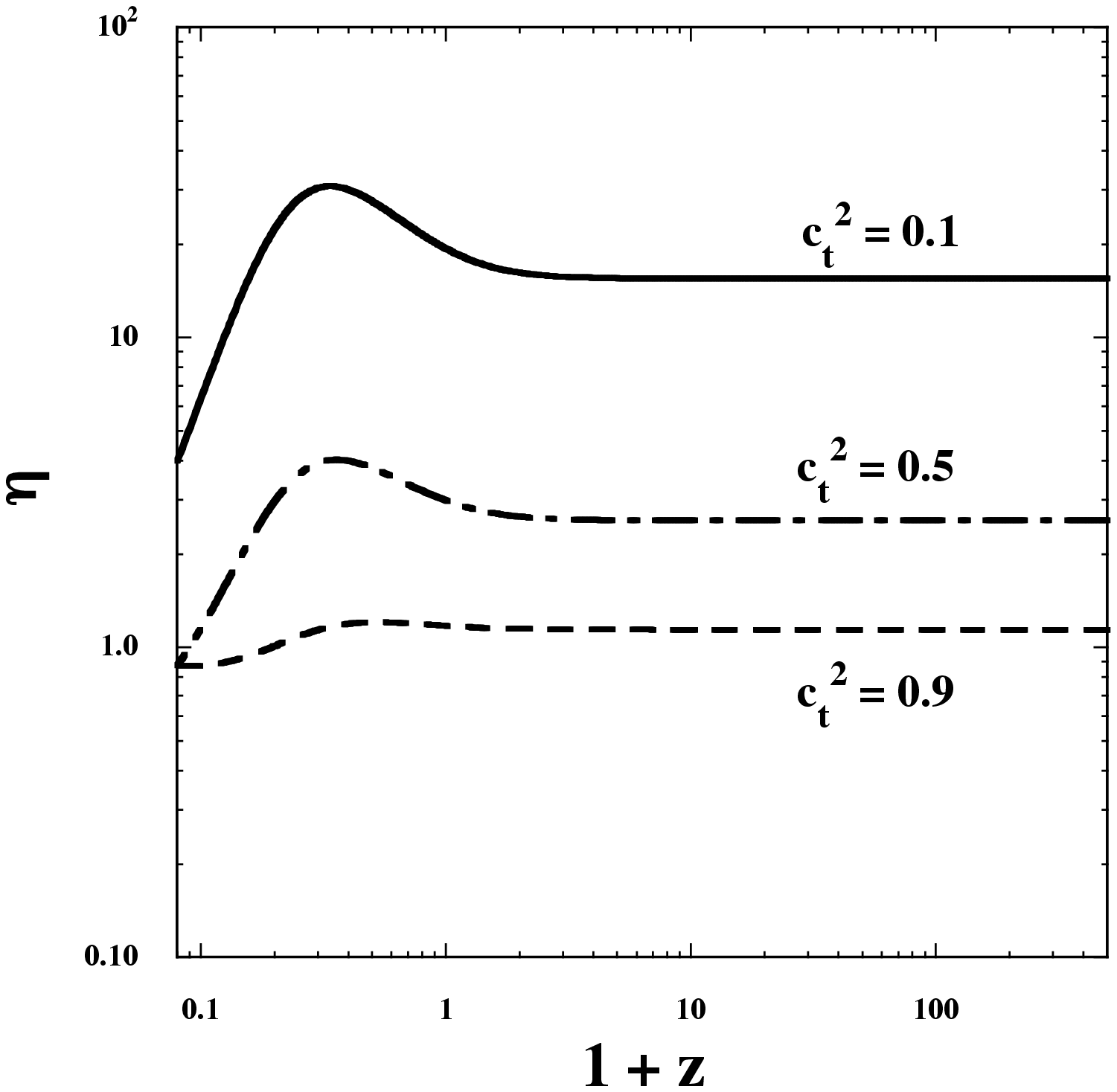}
\caption{
The evolution of the anisotropic parameter $\eta$ versus $1+z$ 
for $c_{\rm t}^2=0.1, 0.5, 0.9$ with the model parameters 
$\lambda_1=10$, $\lambda_2=0$, and $V_2/V_1=10^{-6}$. 
The initial conditions are chosen to be similar to those given 
in Fig.~\ref{fig1}, such that $|V_m^{(h)}| \ll |V_m^{(s)}|$.
\label{fig2}
}
\end{figure}

We confirmed that the analytic solutions (\ref{anaso1})-(\ref{anaso2}) 
show good agreement with the numerical results shown in Fig.~\ref{fig1}. 
In Fig.~\ref{fig2} we plot the evolution of $\eta$ for three 
different values of $c_{\rm t}^2$ smaller than 1.
The anisotropic parameter is nearly constant during the scaling 
matter era. As estimated by Eq.~(\ref{etaana}), the deviation of 
$c_{\rm t}^2$ from 1 leads to the values of $\eta$ larger than 1.
This is the observational signature of our model manifest 
in CMB temperature anisotropies \cite{Planckdark}.
{}From Fig.~\ref{fig1} we find that the two gravitational potentials 
$-\Psi$ and $\Phi$ start to decrease after the onset of the cosmic 
acceleration ($z \lesssim 1$). 
As we see in Fig.~\ref{fig2}, this leads to the variation 
of $\eta$, which signals the end of the scaling matter era.

Substituting the solutions (\ref{anaso1}) into Eq.~(\ref{delms}), 
the perturbation $\delta_m$ during the scaling matter era obeys
the following equation
\be
\delta_m''+\frac12 \delta_m'-\frac32
\frac{G_{\rm eff}}{G}\Omega_m \delta_m \simeq 0\,,
\qquad \quad
G_{\rm eff}=
\frac{K^2(c_{\rm s}^2+1-c_{\rm t}^2)+3\Omega_X}
{c_{\rm s}^2K^2-3\Omega_X}G\,.
\label{delmap}
\ee
As long as $c_{\rm s}^2K^2 \gg \Omega_X$, the effective 
gravitational coupling reduces to 
\be
G_{\rm eff} \simeq \left( 1+\frac{1-c_{\rm t}^2}{c_{\rm s}^2} 
\right)G\,.
\label{Geffap}
\ee
When $c_{\rm t}^2<1$ we have $G_{\rm eff}>G$. 
On using Eq.~(\ref{Geffap}), the growing-mode solution to 
Eq.~(\ref{delmap}) is given by 
\be
\delta_m \propto a^p\,,\qquad
p=\frac14 \sqrt{1+24\Omega_m+24\Omega_m
\frac{1-c_{\rm t}^2}{c_{\rm s}^2}}-\frac14\,.
\label{delmso}
\ee
Compared to the case of GR ($c_{\rm t}^2=1$), the growth 
rate of matter perturbations gets larger for $c_{\rm t}^2<1$. 
However, since $c_{\rm s}^2=c_{\rm t}^2+(\lambda_1^2-3)(1-c_{\rm t}^2)/3$ 
and $\lambda_1 \gtrsim 10$, this modification is suppressed to be small.
For example, even for $c_{\rm t}^2=0$, we have 
$24\Omega_m (1-c_{\rm t}^2)/c_{\rm s}^2=72/\lambda_1^2 \lesssim 0.7$ 
in Eq.~(\ref{delmso}), so the growth index $p$ is not very different from 
that for $c_{\rm t}^2=1$.

\begin{figure}
\includegraphics[width=3.6in]{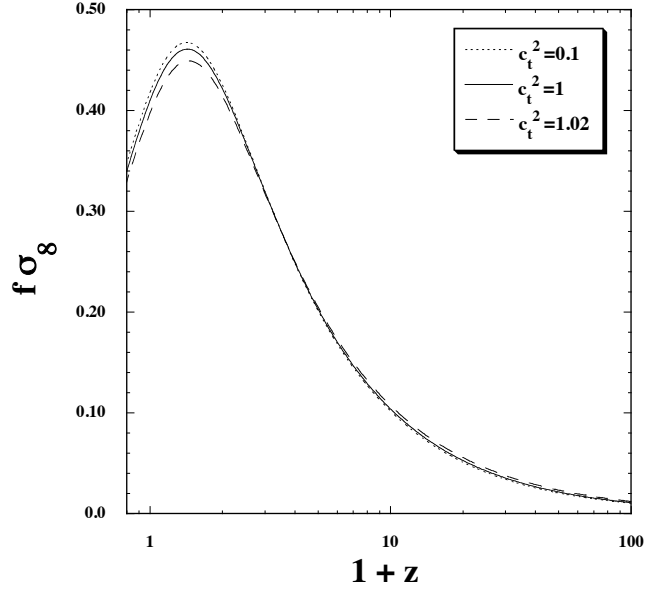}
\caption{The evolution of $f \sigma_8$ versus $1+z$ for 
$c_{\rm t}^2=0.1, 1, 1.02$ with the model parameters 
$\lambda_1=10$, $\lambda_2=0$, and $V_2/V_1=10^{-6}$. 
The initial conditions are chosen to be similar to those given 
in Fig.~\ref{fig1}, such that $|V_m^{(h)}| \ll |V_m^{(s)}|$.
\label{fig3}}
\end{figure}

In Fig.~\ref{fig3} we show the evolution of $f \sigma_8$ for three 
different values of $c_{\rm t}^2$. 
When $c_{\rm t}^2<1$ the effective gravitational coupling 
(\ref{Geffap}) is bigger than $G$, so the growth rate of $\delta_m$ 
tends to be slightly larger than that for $c_{\rm t}^2=1$.
However, the difference between 
the two cases $c_{\rm t}^2=0.1$ and $c_{\rm t}^2=1$
is very tiny except for the low-redshift regime 
in which the cosmic acceleration occurs.

\begin{figure}
\includegraphics[width=3.2in]{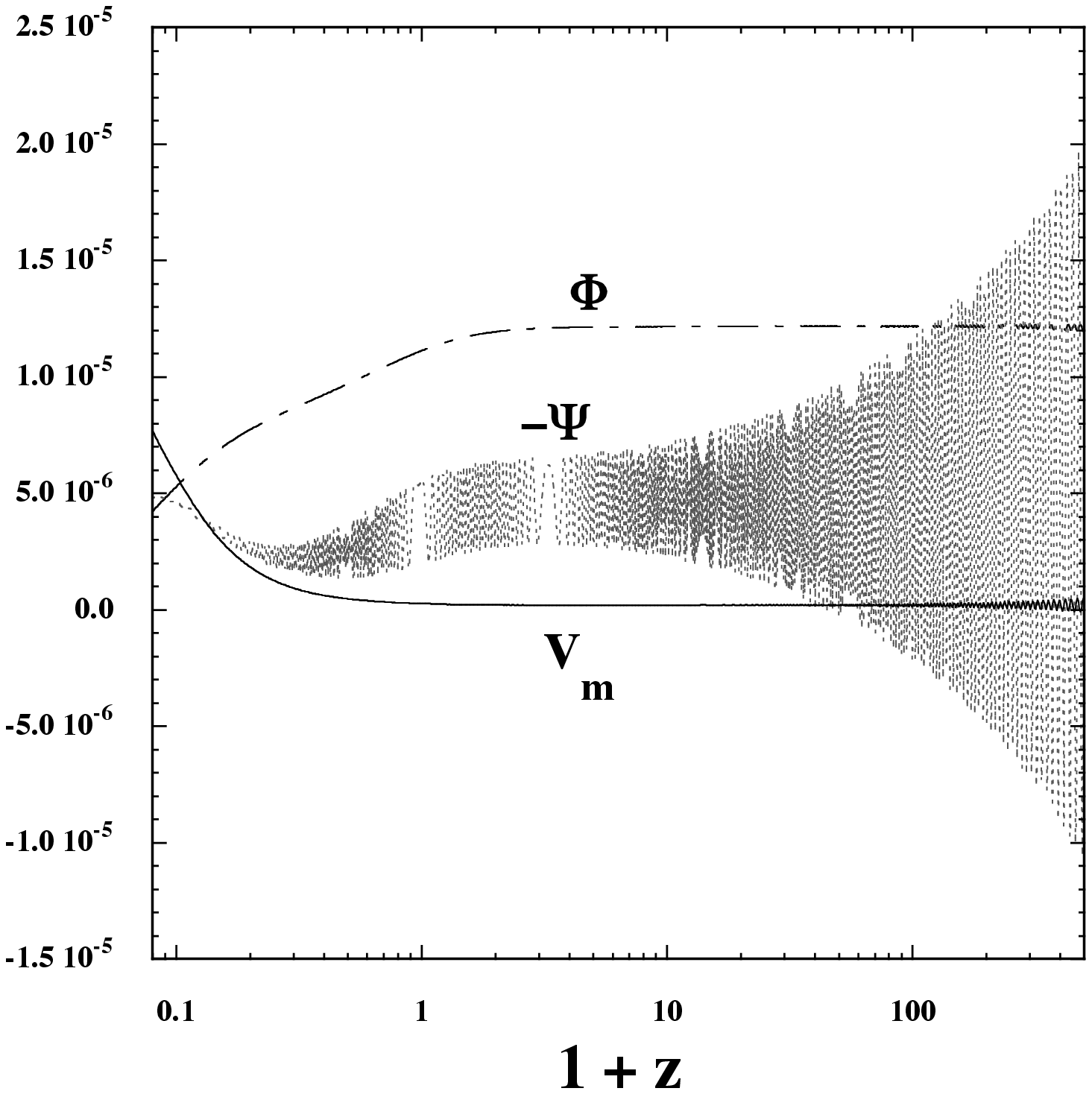}
\caption{The evolution of the velocity potential $V_m$ and the 
gravitational potentials $-\Psi$ and $\Phi$ versus $1+z$
for $c_{\rm t}^2=0.5$, $\lambda_1=10$, 
$\lambda_2=0$, and $V_2/V_1=10^{-6}$
with the initial value $V_m=5.0 \times 10^{-7}$ 
and other initial conditions same as those in Fig.~\ref{fig1}.
\label{fig4}}
\end{figure}

In the super-luminal regime $c_{\rm t}^2>1$, the Laplacian 
instability associated with negative $c_{\rm s}^2$ can be 
avoided for
\be
c_{\rm t}^2-1<\frac{2\Omega_X}{\Omega_m}\,.
\label{ctcon2}
\ee
During the scaling matter era, this condition translates to 
$c_{\rm t}^2-1<3/(\lambda_1^2-3)$.
The dashed curve shown in Fig.~\ref{fig3} ($c_{\rm t}^2=1.02$) 
is the case in which the stability condition (\ref{ctcon2}) is satisfied.
In the low-redshift regime, $f\sigma_8$ for $c_{\rm t}^2=1.02$ 
is slightly smaller than that for $c_{\rm t}^2=1$, so the gravitational force 
tends to be weaker. Numerically we confirmed that, 
for $c_{\rm t}^2$ violating the condition (\ref{ctcon2}), 
the perturbations are prone to violent negative instabilities. 

Finally, we discuss the case in which the initial condition 
$|V_m^{(h)}| \ll |V_m^{(s)}|$ is violated.
The numerical simulation shown in Fig.~\ref{fig4} corresponds to 
such an example, where the initial value of $V_m$ is not close 
to the special solution (\ref{Vms}). 
When $c_{\rm t}^2<1$, the velocity perturbation $V_m$ 
exhibits an oscillation induced by the sound speed 
$c_{\rm s}$ during the scaling matter era.
The amplitude of the homogenous solution decreases as 
$|V_m^{(h)}| \propto a^{-1}$, so $V_m$ approaches 
the special solution (\ref{Vms}) in the end. 
Provided that the approach to $V_m^{(s)}$ occurs 
for $z \gg 1$, the evolution of $f\sigma_8$ in the low-redshift 
regime is similar to that shown in Fig.~\ref{fig3}.

In the numerical simulation of Fig.~\ref{fig4}, we find that 
the gravitational potential $-\Psi$ shows a heavy oscillation 
around $0$ with a large amplitude. 
We recall that $\Psi$ is related to the perturbation 
$\delta N=-V_m'+(H'/H)V_m$, as 
$\Psi=\delta N-(H'/H)\chi+\chi'$.
{}From Eq.~(\ref{Vmh2}) the derivative ${V_{m}^{(h)}}'$ 
can be estimated as 
\be
{V_{m}^{(h)}}' \simeq 
a^{-3/4} \sqrt{\frac{x}{2\pi}}
\left[ -c_1 \sin \left( x-\frac{\pi}{4} \right)+
c_2 \cos \left( x-\frac{\pi}{4} \right) \right]\,,
\ee
where we have assumed $x=2c_{\rm eff}K({\cal N}) \gg 1$.
The amplitude of ${V_{m}^{(h)}}'$  is $x/2$ times as large as
 that of  $V_{m}^{(h)}$.
This is the reason why the amplitude of the oscillating mode of 
$-\Psi$ is much larger than that of $V_m$. 
The definition of the anisotropic parameter 
$\eta=-\Phi/\Psi$ loses its validity whenever $-\Psi$ crosses 0, 
so in such cases we should directly resort to the 
effective gravitational potential (\ref{Phieff}) rather than $\eta$. 
Furthermore, as $-\Psi$ crosses 0, the effective gravitational coupling  
$G_{\mathrm{eff}}$ derived from Eq.~(\ref{Psieq})
changes its sign, leading, in this case, to a rather unclear 
interpretation for the gravitational interaction.
In the numerical simulation of Fig.~\ref{fig4}, the oscillation of 
$-\Psi$ does not damp away even around today ($z \sim 1$).
The amplitude of the oscillating mode of $\Phi=\zeta+\chi$ is smaller 
than that of $-\Psi$, such that the oscillation of the former
disappears in the early stage of the matter era (see Fig.~\ref{fig4}).

The oscillation of the gravitational potential $-\Psi$ leaves 
a distinctive imprint in CMB temperature anisotropies, so it should be 
possible to put tight constraints on the parameter space 
of initial conditions. 
We leave the analysis of observational constraints on the model 
(including the case of initial conditions $|V_{m}^{(h)}| \ll |V_{m}^{(s)}|$) 
for a future work.

\section{Conclusions}
\label{concludesec} 

In the framework of the EFT of modified gravity, we have studied 
the dynamics of cosmological perturbations 
and resulting observational consequences. 
Our starting point is the general action (\ref{geneac}) in 
unitary gauge that involves a gravitational scalar degree of 
freedom and a matter field $\chi$ mimicking a perfect fluid.
This analysis encompasses the GLPV theories 
described by the Lagrangian (\ref{LGLPV}) as a specific case. 

Expanding the action up to second order in the perturbations of 
ADM scalar quantities, we derived the scalar
perturbation equations of motion (\ref{pereq1})-(\ref{pereq4}) 
as well as the equation (\ref{GWeq}) for gravitational waves.
In GLPV theories there is a non-trivial mixing between the 
propagation speeds of matter and gravitational scalar.
The matter sound speed is not strongly affected 
by the deviation from Horndeski theories, but the 
modification to the scalar sound speed $c_{\rm s}$ 
is large enough to be able to distinguish between GLPV and 
Horndeski theories.
For non-relativistic matter with $c_m^2=0$ we derived
the exact relation $c_{\rm s}^2=c_{\rm H}^2-\beta_{\rm H}^2$, 
where $c_{\rm H}$ is given by Eq.~(\ref{csHo}) and 
$\beta_{\rm H}$ is a new term (\ref{betaH}) arising 
in the theories beyond Horndeski.

As an application of our general formalism, we studied the cosmology 
for a simple canonical scalar-field model
described by the Lagrangian (\ref{concrete}).
The deviation from Horndeski theories can be quantified by 
the difference of the tensor propagation speed squared 
$c_{\rm t}^2=F(\phi)$ from 1, where $F(\phi)$ is a function 
of $\phi$ appearing in Eq.~(\ref{AB2}). 
In the regime $|\alpha_{\rm H}|=|c_{\rm t}^2-1| \ll 1$, the scalar sound 
speed squared $c_{\rm s}^2$ is given by Eq.~(\ref{cses}).
For dark energy models in which the ratio $\Omega_m/\Omega_X$ 
gets larger in the past (which is the case for most of models), 
$|c_{\rm s}^2|$ can be much larger than 1 
even for $|\alpha_{\rm H}| \ll 1$. 
In particular, for $c_{\rm t}^2>1$, the scalar 
perturbation is plagued by the Laplacian instability 
unless the condition (\ref{ctcon}) is satisfied.

The behavior of large values of $c_{\rm s}^2$ in the past can be 
avoided for the dark energy model given by the potential  
(\ref{poten}), in which case there exists the scaling solution 
characterized by the constant ratio
$\Omega_m/\Omega_X=2(\lambda_1^2-3)/3$. 
In this model we studied the evolution of cosmological perturbations 
to find observational signatures in the framework of GLPV theories.
The solution to Eq.~(\ref{Vmeq}) for the velocity potential 
$V_m=Hv_m$ can be written in terms of the sum of 
the special solution $V_m^{(s)}$ and the homogenous 
solution $V_m^{(h)}$. During the scaling matter era 
$V_m^{(s)}$ stays nearly constant, whereas 
$V_m^{(h)}$ decreases in proportion to $a^{-1}$ 
with oscillations.

Provided that $|V_m^{(h)}| \ll |V_m^{(s)}|$, 
the evolution of perturbations during the scaling 
matter era can be known analytically. Of course this  
requires the fine-tuning of initial conditions.  
In this case, the anisotropic parameter 
$\eta=-\Phi/\Psi$ between the two gravitational potentials is 
expressed in terms of $c_{\rm t}^2$ and $c_{\rm s}^2$ alone. 
As we see in Fig.~\ref{fig2}, the deviation of $\eta$ from 1 tends to 
be larger as $c_{\rm t}^2$ is away from 1. 
We have also estimated the growth rate of matter perturbations 
$\delta_m$ and found that the evolution of $\delta_m$ is not 
sensitive to the tensor propagation speed ranging 
in the regime (\ref{ctcon2}).

If $V_m^{(h)}$ dominates over $V_m^{(s)}$ at the initial stage of 
the matter era, the perturbations exhibit rapid oscillations with 
the frequency related to $c_{\rm s}^2$ until $V_m$ approaches 
the special solution $V_m^{(s)}$. 
The amplitude of oscillations is particularly large for the gravitational 
potential $\Psi$, see Fig.~\ref{fig4}.
This property comes from the fact that $\Psi$ is related to 
the derivative term ${V_m^{(h)}}'$, 
whose amplitude is $x/2$ times 
as large as that of $V_m^{(h)}$.

We have thus shown that the simple extension of Horndeski theories 
to GLPV theories gives rise to interesting observational signatures. 
For the initial conditions satisfying $|V_m^{(h)}| \ll |V_m^{(s)}|$, the anisotropic 
parameter $\eta$ deviates from 1 even in the early stage of the 
matter era (which is usually not the case in Horndeski theories).
For other initial conditions, the gravitational potential $\Psi$ 
rapidly oscillates with a large amplitude especially when 
$c_{\rm t}^2$ is away from 1. 
In both cases, we should be able to put tight bounds on the 
deviation parameter $\alpha_{\rm H}=c_{\rm t}^2-1$ from the observations of 
CMB and weak lensing. It will also of interest to study local gravity 
constraints on our model along the lines of Refs.~\cite{KWY,Koyama:2015oma,Saito}.

\section*{ACKNOWLEDGEMENTS}
KK is supported by the UK Science and Technology Facilities Council (STFC) 
grants ST/K00090/1 and ST/L005573/1. 
ST is supported by the Scientific Research Fund of the JSPS (No.~24540286) 
and by the cooperation programs of Tokyo University of Science and CSIC. 
We thank the organizers of the ``Testing Gravity 2015'' workshop 
held in Vancouver at which this work was initiated.


\appendix
\section{Quasi-static approximation on sub-horizon scales 
in Horndeski theories}
\label{secappen} 

In this Appendix we derive the expression of $G_{\rm eff}$ and 
$\Phi_{\rm eff}$ in Horndeski theories under the quasi-static 
approximation for the modes deep inside the sound horizon.
Since $\alpha_{\rm H}=0$ in this case, we can solve 
Eqs.~(\ref{quasi2}) and (\ref{quasi3}) for $\psi$ and $\zeta$, as
\be
\psi=\frac{L_{,\cal S}(4b_1-c_{\rm t}^2{\cal W})}
{b_2 c_{\rm t}^2 L_{,\cal S}-4b_1^2}\Psi\,,\qquad
\zeta=\frac{b_1{\cal W}-b_2L_{,\cal S}}
{b_2 c_{\rm t}^2 L_{,\cal S}-4b_1^2}\Psi\,,
\label{psizeta}
\ee
where 
\be
b_1 \equiv \dot{L}_{,\cal S}+HL_{,\cal S}\,,\qquad
b_2 \equiv \dot{\cal W}+H{\cal W}+\rho\,.
\ee
Substituting the relations (\ref{psizeta}) into Eq.~(\ref{quasi1}) 
with the approximation $\delta_m \simeq \delta$, we obtain 
Eq.~(\ref{Psieq}) with the effective gravitational coupling 
\be
G_{\rm eff}=\frac{b_2 c_{\rm t}^2 L_{,\cal S}-4b_1^2}
{4\pi L_{,\cal S}({\cal W}^2 c_{\rm t}^2 +
4b_2L_{,\cal S}-8b_1{\cal W})}\,.
\label{Geff2}
\ee
{}From Eq.~(\ref{psizeta}) the gravitational potential 
$\Phi=\zeta+H\psi$ satisfies the relation $\Phi=-\eta \Psi$, 
with the anisotropic parameter
\be
\eta
=\frac{b_2 L_{,\cal S}
-b_1{\cal W}+HL_{,\cal S}({\cal W}c_{\rm t}^2-4b_1)}
{b_2 c_{\rm t}^2 L_{,\cal S}-4b_1^2}\,.
\label{eta2}
\ee
On using Eq.~(\ref{Psieq}), (\ref{Geff2}), and (\ref{eta2}), the effective 
gravitational potential (\ref{Phieff}) obeys
\be
\Phi_{\rm eff}=
-\frac{a^2}{k^2} \frac{b_2 (1+c_{\rm t}^2)L_{,\cal S}
-4b_1^2-b_1{\cal W}+HL_{,\cal S}({\cal W}c_{\rm t}^2-4b_1)}
{2L_{,\cal S}({\cal W}^2 c_{\rm t}^2 +
4b_2L_{,\cal S}-8b_1{\cal W})}
\rho \delta_m\,.
\label{Phieff2}
\ee
The results (\ref{Geff2})-(\ref{Phieff2}) match with those derived 
in Ref.~\cite{DKT} in the Newtonian gauge 
by taking the massless limit $m_{\phi} \to 0$.


\end{document}